\documentclass[traditabstract]{aa}
\usepackage{graphics}
\usepackage{txfonts}
\usepackage{bm}
 \usepackage{graphicx}
\usepackage{natbib}
\bibpunct{(}{)}{;}{a}{}{,} 

\begin{document}
\def\be{\begin{equation}}
\def\ee{\end{equation}}
\title{Solving the transport equation by the use of
  6D spectral methods in spherical coordinates}
\subtitle{}

\author{S.~Bonazzola\inst{\ref{Lut}}
,
  N.~Vasset\inst{\ref{Bas}, \ref{Lut}}
  \and
  B.~Peres\inst{\ref{Lut}}}
\institute{Laboratoire Univers et Th\'eories, Observatoire de Paris, CNRS,
  Universit\'e Paris Diderot, 5 place Jules Janssen, F-92190, Meudon,
  France.\\
  \label{Lut}
  \and
  Universitaet Basel, Departement Physik, Klingelbergstrasse 82, CH-4056 Basel, Switzerland.\\
  \email{silvano.bonazzola@obspm.fr, nicolas.vasset@unibas.ch, bruno.peres@obspm.fr}\label{Bas}
}

\date{ 6 June 2011 / 6 July 2011}

\abstract{ We present a numerical method for handling the resolution
  of a general transport equation for radiative particles, aimed at
  physical problems with a doubly spherical geometry. Having in mind
  the computational time difficulties encountered in problems such as
  neutrino transport in astrophysical supernovae, we propose a scheme
  based on full spectral methods in 6d spherical coordinates. This
  approach, known to be suited when the characteristic length of the
  dynamics is much smaller than the domain size, has the potential
  advantage of a global speedup with respect to usual finite
  difference schemes. An analysis of the properties of the Liouville
  operator expressed in our coordinates is necessary in order to
  handle correctly the numerical behaviour of the solution. This
  reflects on a specific (spherical) geometry of the computational
  domain.  The numerical tests, performed under several different
  regimes for the equation, prove the robustness of the scheme: their
  performances also point out to the suitability of such an approach
  to large scale computations involving transport physics for massless
  radiative particles. {\bf We wish to point out that the algorithm
    presented here, is particularly suitable to treat problems in
    which matter has high velocities, such as the neutrino transport
    in supernovae.}}  \keywords{ Transport equations -- Radiative
  astrophysics -- Spectral methods }

\titlerunning{Solving the transport equation with 6D spectral methods}
\authorrunning{S.~Bonazzola, N.~Vasset \& B.~Peres}
\maketitle

\section{Introduction} 
Particle transport phenomena are central in modelling systems governed
by radiative hydrodynamics, encountered very often in astrophysics as
well as in plasma physics. A global description of radiative transport
involves the hyperbolic transport equation, sometimes called Boltzmann
equation in the literature\footnote{ The main difference between the
  two concepts is that in the transport equation, the collision term
  only describes interactions between neutrinos/photons and external
  medium (atoms, nuclei or electrons). (Herein after, we shall refer
  to neutrinos and photons as ``radiative particles'', only using the
  term photon or neutrino when it turns out to be necessary).  In the
  Boltzmann case, the collision terms also describe in principle
  interactions between the radiating particles; these are not relevant
  in our context. Therefore, we will try to avoid the term ``Boltzmann
  equation'' from now on.}.  This equation describes the time
evolution of a distribution function $F$ defined on a 6-dimensional
phase space.
The high dimensionality of this equation often prevents its numerical
resolution in the most general geometry, due to unaffordable
computational resources for obtaining physical results in a reasonable
CPU time, when using classical techniques. As a result, most numerical
models for radiative transport physics in several settings either
restrict the global geometry of the problem (as in~\cite{Mezza1989,
  Gourg1993, Lie2005, Muel2010}), or replace the transport equation by
simplified models usually involving the distribution moments
(~\cite{Anders1972, Lev1979, Thorne1981, Carda2003,
  Lie2009}). However, in the particular setting of neutrino transport
in astrophysical supernovae, the fact that a multidimensional
transport model for neutrino radiation is required to reproduce the
observed supernova explosions has been strongly hinted in recent
simulations~\citep{Lie2005}. As a result, attempts have been made in
this direction, but they result in very demanding simulations, that
are still not able to capture all the needed physics in a general
geometry~\citep{Mess2008, Mare2009}.

Apart from the problem of
 dimensionality and size of the simulations, transport phenomena very often 
involve physical processes occurring through several orders of
 magnitude 
for typical lengths. Once again, this problem arises 
in the supernovae neutrinos setting, when comparing the mean free path
 of a 
radiative particle in the diffusion regime and the typical size of the system~\citep{Jan2007}.  
As long as the two computational problems mentioned above are
concerned, 
it is customary to privilege numerical methods with a high order of
accuracy~\citep{LeV2002}. 
The use of multidimensional spectral methods~\citep{Gott1977,Canu1988,Canu2006} seems to be
 especially adequate.
 
Concerning the computational size difficulties, a rule of thumb~\citep{Gott1977}
 claims in fact that for a given accuracy, spectral treatment
  requires five times less grid points per dimension then the ordinary second order
  finite difference algorithm. A computational factor of $5^6=15625$ 
 could then be gained in modelling the transport equation. Consequently, solving the 6-D 
transport problem in a reasonable time while using a spectral algorithm and massive parallelisation seems possible.
In this article, we shall describe the numerical methods 
that can be used in solving the sole transport equation for radiative particles
 and testing such an economy in computational time. 

 From a mathematical 
point of view and adopting suitable approximations, the neutrino
and photon transport equations are analogous in a wide range
 of settings. The abundant results on the photon transport equation will be however
 mainly used here for testing numerically our scheme\footnote {The
   only difference between general 
transport equation for neutrinos and photons, apart from a difference
in cross section expressions, 
is the sign in front of the non-linear terms accounting for induced processes.}.

In this work, the following assumptions are made:

1) As mentioned above, we assume the mass of the radiative particles to be zero
(which for astrophysical neutrinos is a fairly 
reasonable assumption).

2) Neutrino and photon polarisation states are not taken into
  account, and are averaged on.

3) The interacting plasma is assumed to be in local thermal
equilibrium. This physical oversimplification will allow for a much simpler 
treatment, by enabling to introduce the
full thermal equilibrium limit in 
the equations. 

4) For the simplicity of discussion, possible general relativistic
terms are not taken into account. We believe that this aspect,
though physically important in some astrophysical computations, will
in no way change the behaviour of the numerical scheme, or the
mathematical properties of the studied equations.
  
Using (6+1) general spherical coordinates in phase space, we will
perform a numerical resolution using spectral methods based on
Fourier/Chebychev expansions, depending on the type of coordinates
involved. This expansions will be performed in the physical space
(classical spherical coordinates $(r, \theta, \phi)$) as well as in
the momentum space (energy dependence and angular coordinates for the
momentum part). Those methods have been developed and extensively used
by our group and were described for the first time in~\cite{Bona1985}
(see also the review of~\cite{Gran2009}).  Due to possible
discontinuities arising on the space variable $r$, we will also
propose a hybrid version of the code in which we use finite
differences only in this dimension. A tentative conservative full
spectral version including also a Chebychev decomposition in $r$ will
be designed and tested as well.  We shall show in different contexts
relevant results in 5 dimensions at most, and give the corresponding
CPU time obtained for every simulation, on single-processor runs of an
ordinary computer with a clock frequency of $2.5$ Ghz.

The paper is organised as follows: in Sect.~\ref{sect:mathframework},
we present the general mathematical framework of the transport
equation, alongside with typical physically motivated source terms for
the equation.  In Sect.~\ref{sect:diffapprox} we present two
asymptotic settings for physical transport of neutrinos and photons,
namely the coherent transport case and the Fokker-Planck
approximation. Those two limiting cases will be used as test problems
in the ensuing numerical investigations.  Sect.~\ref{sect:sphcoord}
will present the derivation of the transport equation in our chosen
coordinates, as well as the possible mathematical problems that arise
with this description; a few solutions will then be proposed to handle
resolution in the most efficient way.  Sect.~\ref{sect:numtest}
presents first numerical tests of the designed code, including time
evolution for uniform distribution and full coherent transport using
hybrid discretization. In Sect.~\ref{sect:rotns} we show an
application of the method to the transport of neutrino in a rotating
neutron star by using an diffusion approximation in the inner part of
the star and the exact solution
in the outer region where the the diffusive approximation fails.

Appendix~\ref{app:partcons} presents an example of an explicitly
particle-conservative form of the solved equations.
Appendix~\ref{app:consform} presents a solution to the full spectral
approach in spherical symmetry, with an emphasis on the issue of
conservation of the number of particles. {\bf In
  Appendix~\ref{app:diffeq} we show how the diffusion equation and
  telegraph equation are obtained. In our opinion, the telegraph
  equation is more suitable to treat neutrino transport when the
  matter velocity is close to the light velocity (supernovae
  problems).  In fact the solution of the telegraph equation are such
  that the propagation velocity of the neutrino is always $ \le c /
  \sqrt 3$ .  We want to point out that the telegraph equation can be
  numerically implemented with minor modifications with respect to the
  diffusion equation. }
\section{Transport equation in 6 dimensions}
\label{sect:mathframework}

\subsection{Context and definitions}

 Let $f(x,y,z,p_x,p_y,p_z,t) $ be the distribution function in the 
phase space for  a collection of particles, expressed in Cartesian-like coordinates. The transport equation
will quantify the evolution of this distribution function with respect to collision terms, that
describe the interaction of the radiative 
particle with other particle species of a plasma. The change in number
of particles in the elementary phase space volume 
  $D^3 \vec{x} \, D^3 \vec{p}=dx\, dy\, dz\, dp_x\,dp_y\, dp_z $ is then described by
\be\label{Boltz1} 
  \frac{D}{dt} f D^3 \vec{x} \, D^3 \vec{p} =CT,
\ee
 where $D/dt$ represents the total derivative and $CT$ includes the collision terms. A reasonable assumption for
neutrinos or photons is that between 
collisions with plasma particles (described in the collision terms),
radiating photons/neutrinos travel in straight 
lines with no change in energy. This amounts to the absence of global forces acting on the radiating particles.
 In this context, the non-general relativistic transport equation in Cartesian coordinates takes the form:
\be\label{Tra1} 
\frac{1}{c}\frac{\partial f}{\partial t}+\omega^i \frac{\partial f}{\partial
 x^i}= CT,  
\ee
where
\be\label{betai}
\omega^i= \frac{p^{i}}{||p||}, \;\;\;\; \vec{\omega} \cdot \vec{\omega}=1,
\ee
$c$ is the light velocity, and $CT$ is a collision term that depends also on $f$.
\subsection{The collision terms}
 Three different types of processes are expressed in collision terms:
\begin{itemize}
\item The rate of spontaneous emission of radiating particles by a
  particular process in the plasma will 
be expressed as $S(\nu,\vec{x})$; here $ \vec{x}$ models the spatial
  position, and $\nu$ is the radiative particle frequency
  (or its reduced energy $E/h$, where $h$ is the Planck constant).
 A  general assumption is that the emission is 
isotropic (i.e. matter itself does not have a preferred
  direction). This of course is only valid 
if our frame of reference is moving with the plasma.
\item Absorption processes are expressed by a cross section
  $\sigma_{a}(\nu,\vec{x})$ and are also assumed 
to be isotropic.
\item Scattering processes at a spatial position $\vec{x}$, from a
  radiating particle scattered 
from coordinates $(\vec{\omega},\nu)$ within $(d\vec{\omega},d\nu)  $
  to $(\vec{\omega}',\nu')  $ 
within $(d\vec{\omega}',d\nu')$ is expressed by the cross section 
$\sigma_d \left(\vec{x},\vec{\omega} \cdot \vec{\omega^{'}}  ,\nu \to
  \nu^{'} \right) $. Again, in agreement 
with the previous assumption of isotropy for matter processes, this
  differential cross section only 
depends on the angle between the incoming and scattered radiating
  particle momentum, 
{\it via} the simple scalar product $\vec{\omega} \cdot \vec{\omega}^{'}$. 
\end{itemize}

 We further assume for simplicity that radiative particles interact
 with only one species 
of the plasma, of particle mass $m_{0}$. Using this energy scale we define an auxiliary distribution function 
\be
  F= \left( \frac{h\nu}{m_{0}c^{2}} \right)^{2}f = \gamma^{2}f,
\ee
where the notation $\gamma$ is coined as the dimensionless energy for
 the radiative particle. 
This redefinition of the distribution function allows for a slightly
 simpler notation 
for the interaction terms, while the left hand side operator of Eq.~\ref{Tra1} keeps the same form as applied to $F$. 
Once all those quantities characterising the interactions with matter are known, 
the collision term is determined by the formula:\\
\\
\begin{eqnarray}
\label{Transp}
CT= n(\vec{r},t) \{ S(\nu) -\sigma_a(\nu)F  \nonumber \\
+\int_0^{\infty} d \nu^{'} \int_{4 \pi} d \vec{\omega}^{'}
\sigma_d({\bf{\nu^{'} \to \nu}},\vec{\omega} \cdot \vec{\omega^{'}} )F(\nu^{'},\vec{\omega^{'}})   
[1 \pm \frac{c^3}{2 \nu^2} F(\nu, \vec{\omega})] \nonumber \\
-\int_0^{\infty} d \nu^{'} \int_ {4\pi} d\vec{\omega}^{'} \sigma_{d}(\nu
\to \nu^{'}, \vec{\omega} \cdot \vec{\omega}^{'} ) F(\nu,\vec{\omega})
[1 \pm \frac{c^3}{2 \nu^2} F(\nu^{'} , \vec{\omega}{'})] \},  
\end{eqnarray}
where $n(\vec{r},t) $ is the interacting plasma density. For clarity, no spatial dependence of the chemical composition for the plasma is assumed; we finally write here interaction terms that are only proportional to the plasma density $n$. In the
right hand side part, 
and besides the absorption and emission terms, 
the first integral term  models the in-scattered neutrinos to
coordinates to 
$(\vec{\omega},\nu) $ within $(d\vec{\omega},d\nu)$. The second
  integral term models the 
out-scattered neutrinos, from $(\vec{\omega},\nu) $ to
  $(\vec{\omega}',\nu')$. $\sigma_d$ is the differential scattering kernel for interactions.
 We have also included in front of the scattering integrals the
 quantum corrections due to induced processes for both types of
 radiating particles\footnote{ The plus sign holds for
     bosons (photons), the minus sign 
holds for  fermions (neutrinos)}. Only one
 term for each type of process is represented
in an attempt for concision.

\section{Different approximations} 
\label{sect:diffapprox}

\subsection{ The coherent scattering}
 Consider the very low energy regime  for the plasma and the radiating
 particles; 
the following assumptions are then made:

1) The plasma is at rest in our frame.

2) The ratio between the scattered particle energy $ h \nu$ and
  the scattering 
target rest mass $ m_{0}c^{2}$ (be it a lepton or a hadron) is very small: 
($\gamma = h \nu/ m_{0}c^2 << 1 $).The velocity of scattering
  plasma particles will then always be neglected in this case.\\
Under the above assumptions, we crudely approximate 
that no energy exchange occurs, meaning that the energy of the
scattered particle is the same 
as the incoming one. Therefore the scattering kernels writes:
\be\label{sigmadif}
\sigma_d(\nu \to \nu^{'},\vec{\omega} \cdot \vec{\omega}^{'})=
 \sigma_d (\vec{\omega} \cdot \vec{\omega}^{'}) \delta(\nu-\nu{'}),
\ee 
where $\delta$ is the Dirac function. Here we give general expressions
for the differential and total 
cross sections $\sigma_d$ and $\sigma_t$, for photon and neutrino
scattering to electrons and hadrons. 
The total cross section is defined by:
\be\label{sigmatot}
\sigma_t= \int_0^{\infty} d \nu^{'} \int_{4 \pi} d { \omega}^{'}
\sigma_d(\nu \to \nu^{'}, \vec{\omega} \cdot \vec{\omega}^{'} ).
\ee

 In the photon/electron case, the coherent scattering approximation
 leads to the well-known Thomson scattering cross sections:
\be\label{sigmaTh}
\sigma^{Th}_d=2 r_e^2 (1+(\vec{\omega} \cdot \vec{\omega}^{'})^2) \,
\delta( \nu-\nu^{'}),   \; \; \; \sigma^{Th}_t=\frac{8 \pi}{3} \, r^2_e 
\ee
where $r_e= e^2/(m_e c^2) $ is the classical radius of the electron,  $m_e$ being its mass.

 In the neutrino/hadron interaction case, the differential cross
 section is usually reduced to the two 
leading orders in the angular decomposition, in the form: 
\be\label{sigman}
\sigma^n_d =\frac{1}{4 \pi} (A +B (\vec{\omega} \cdot
\vec{\omega}^{'})) \, \delta(\nu-\nu^{'})
, \; \; \; \sigma^n_t =A 
\ee
where $A$ and $B$ are constants depending on weak interaction parameters.

\subsection{The Fokker Planck approximation}
 For a plasma particle, we denote by $\alpha=kT/m_{0}c^2$ 
the thermal energy to mass energy ratio. In this section 
we assume that $\gamma << 1$ as before, $\alpha << 1$ 
and that the plasma, at rest in the laboratory frame, is in local
thermodynamic equilibrium. 
In this context, we would like to describe low order energy
redistribution in scattering processes.
  This is the setting of the Fokker-Planck approximation \footnote{This approximation holds for photon
nucleon collisions at plasma temperatures $ < 10^{11} \ \textrm{K} $ and for temperature $T \le 10^9 K^0$  and photon energy
$ h \nu \leq 0.1 \, \textrm{MeV} $. It is especially relevant in the context of X-ray
astrophysics}. We obtain then for the photon distribution 
function $F=\gamma^{2}f$~ (\cite{Pomr1973} \ Eq.(8.62)) :       
\begin{eqnarray}
\frac{\partial F(\gamma \vec{\omega})}{\partial t}+ \vec{\omega} \cdot
\nabla F(\gamma,\vec{\omega})=n(\vec{x},t) \sigma_{Th} 
\left\{ -F(\gamma,\vec{\omega}) \right. \nonumber \\
\left. +\frac{3}{16 \pi} \int_{4 \pi} d \vec{\omega}^{'}  
\left[1+(\vec{\omega} \cdot \vec{\omega}^{'})^2 \right]
F(\gamma,\vec{\omega}^{'} ) \right.
\nonumber \\
\left. +\frac{1}{4 \pi}\frac{\partial}{\partial
  \gamma}  
\left( \alpha \gamma^2 \frac{\partial}{\partial_\gamma} +\gamma^2 -2 \alpha \gamma\right) \int_{4 \pi} d \vec{ \omega}^{'}
F(\gamma,\vec{\omega}^{'}) \right. \nonumber \\
\left. - \frac{3}{128 \pi^2}
 \int_{4 \pi} d \vec{\omega}^{'} 
\int_{4 \pi} d \omega^{``} \left[ 1-\vec{\omega}^{'} \cdot \vec{\omega}^{``} 
+(\vec{\omega}^{'} \cdot \vec{\omega}^{``})^2 -(\vec{\omega}^{'} \cdot
\vec{\omega}^{``})^3 \right] \right. \nonumber \\
\left. \partial_\gamma
\left(F(\gamma,\vec{\omega}^{'} F(\gamma, \vec{\omega}^{``} ) \right) \label{Transpfp}
\right\},
\end{eqnarray}
where we used here a number distribution function, as opposed to the
energy distribution function $I$  in~\cite{Pomr1973}. 
The equation is written in the reduced length unit of
$\lambda_c=h/m_{0}c $ \
, $\lambda_c =2.42^{-10} $ cm being the Compton wavelength.
As before, some variables dependencies in the distribution function are implicit.
Absorption and emission terms are also not written here.\\

By using the Fokker Planck approximation, one can then replace
 the integral operator on the energy in the Eq.(\ref{Transpfp})
by a differential operator, much easier to handle numerically.
As mentioned in the introduction, the Fokker-Planck limit 
for transport will be considered as a test case in numerical 
investigations, alongside with the coherent scattering limit.
                        
From the Fokker-Planck equation we can define two typical times: the ``isotropisation time'' 
$\tau_{is}=1/(n \,\sigma_{Th} \, c) $ describes the typical 
evolution of the angular distribution in phase space, whereas the ``bosonisation time''
$\tau_{Bo}=1/(\alpha \,n \sigma_{Th} \,c)$ 
will be related to dynamical changes in the energy spectrum of
radiative particles. 
Since $\alpha << 1 $, $ \tau_{Bo} >> \tau_{is} $ holds;  
consequently, during the evolution, $F$ will undergo an
``isotropisation'' process in a shorter timescale than the
energy spectrum of the distribution 
function $F$ will change significantly. 
If $F$ is homogeneous and depends only on $t$ and $\gamma$, then
Eq.(\ref{Transpfp}) reduces 
after integration on $\vec{\omega}^{'}  $ and
$\vec{\omega}^{``} $  to the Kompaneet equation ~\citep{Komp1957}:
\be\label{Kompa}
\frac{1}{c} \frac{\partial F}{\partial \,t} + n \, \sigma_{Th} \frac{\partial  }{\partial_\gamma}
\left[ \alpha \gamma^2 \frac{\partial F}{\partial \gamma} 
+ (\gamma^2-2 \alpha \gamma) F + \frac{1}{2} F^2 \right]=0.
\ee

If we integrate both sides of the
Eq.(\ref{Kompa}) on the dimensionless energy $\gamma$, we obtain, as
expected, an equation which expresses the conservation of the number
of photons:
\be\label{Cosph}
\frac{\partial}{\partial t} \int_0^{\infty} F(\gamma,t) d \gamma =0.
\ee

 The steady state solution of Eq.(\ref{Kompa})
is a Bose distribution:
\be\label{Bose}
F(\gamma)= \gamma^{2}f(\gamma) =
2 \gamma^2 \left(\exp{(\frac{\gamma-\mu}{\alpha})} - 1 \right)^{-1},
\ee
The factor of 2  in the right hand side being related to photon
  polarisation. $\mu $ 
 is an integration constant which physically represents the chemical potential.

 The steady state solution is then a Bose distribution and {\it not}  the usual 
Planck distribution that describes full thermal equilibrium. This is due to the fact that we have
omitted the absorption and emission terms, and consequently 
constrain the photon number conservation given by the Eq.(\ref{Cosph}). 
This also justifies the term of ``bosonisation'' introduced
above.
\section{The transport equation in spherical coordinates} 
\label{sect:sphcoord}

\subsection{Definitions and properties}

\begin{figure}
\center
\includegraphics[width=0.6\textwidth]{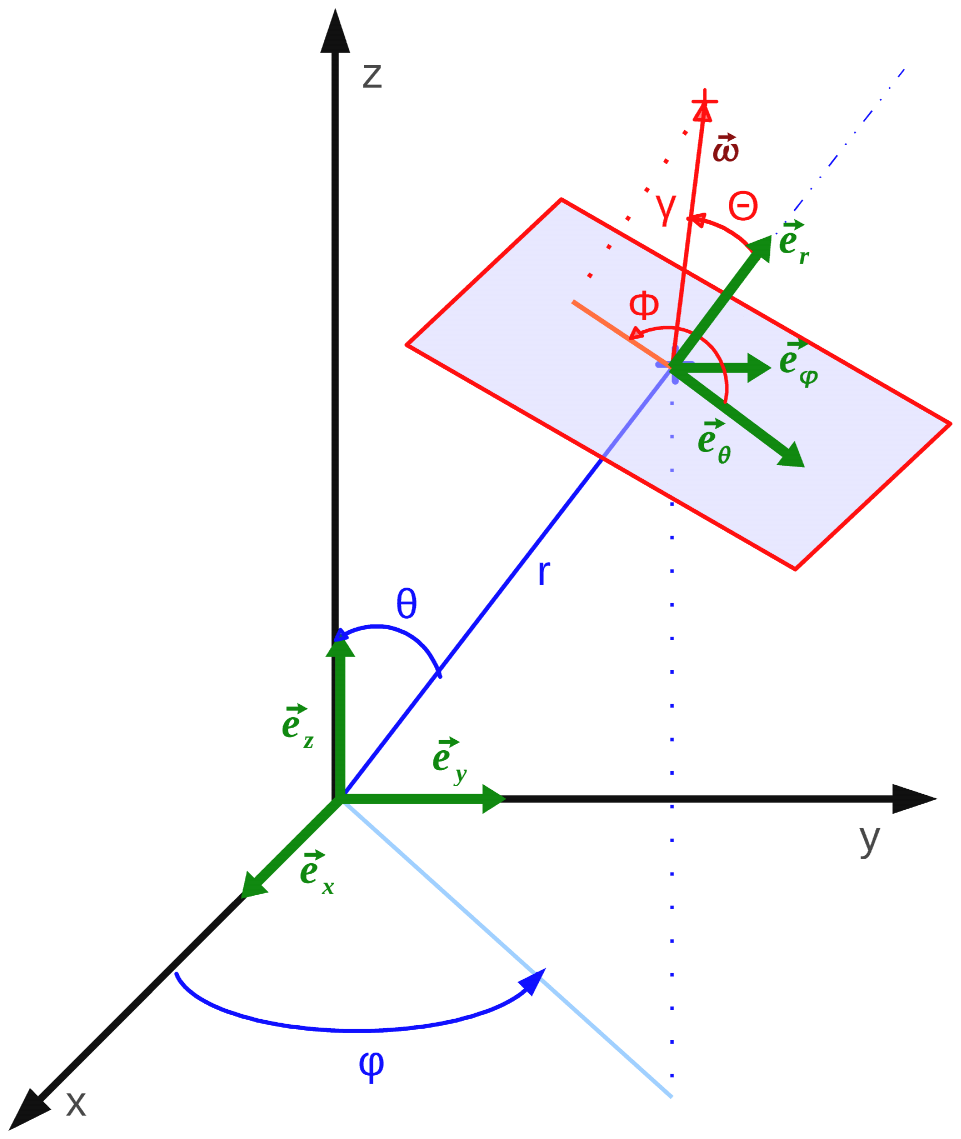}
\caption{Schematic representation of the 6d spherical coordinate in
  phase space. Note that a 
freedom exist for choosing the angular variable $\Phi$ up to a constant.}
\label{f:Sph6d}
\end{figure}

 Starting from the quite general expression for the above equations, we
specify now the geometry of our 
setting, as well as the attached chosen system of coordinates. Having
in mind transport modelling in astrophysical (stellar) settings, 
the most natural geometry for this type of study is the
spherical one. We here choose a set of 
6-D spherical coordinates related to previously defined phase space vectors
$(\vec{r}, \vec{\omega})$,
 and described by the variables $r, \theta, \varphi, \gamma, \Theta, \Phi$ as in
Fig.~\ref{f:Sph6d}. 

 The first three variables are the 
classical 3D spherical coordinates in physical space; $\Theta$ and
$\Phi$ represent the angular 
dependence in the momentum space, whereas $\gamma $ is a
dimensionless measure of the photon (resp. neutrino) energy. 
In this system of coordinates, we can, from the expression in 6-D
Cartesian-like coordinates, write the Liouville operator $\vec{\omega}
\cdot \vec{\nabla} = {\cal L} $ using Jacobi matrix products for
coordinate changes; one has however to keep in mind that in the new
coordinate set, the angular variables in the momentum space are
defined with respect to physical space angular coordinates; this of
course slightly complicates the calculation.  In the end, the operator
${\cal{L}}_{sph}$ (in doubly spherical coordinates)
reads(~\cite{Pomr1973},and references therein):
\begin{eqnarray}
 \label{Liouvsph}
{\cal{L}}_{sph}= \cos \Theta \, \frac{\partial}{\partial r} 
+\frac{1}{r} \left[ \sin \Theta \cos \Phi \, \frac{\partial}
{\partial \theta} \right. \nonumber \\
\left. +\frac{\sin \Theta \sin \Phi}{\sin \theta} \, \frac{\partial}{\partial \phi}
-\sin{\Theta} \, \frac{\partial}{\partial \Theta}  
-\sin \Theta \sin \Phi \frac{\cos \theta}{\sin \theta} \,
\frac{\partial}{\partial \Phi} \right], 
\end{eqnarray}
so that the general transport equation becomes
\be\label{Transpfpsph}
\frac{1}{c} \frac{\partial F}{\partial t}+ 
{\cal L}_{sph} \, F = CT.
\ee

Let us note that it is also possible (and useful) to write this
 equation in a conservative form: see the Appendix A for a derivation of it. 

In order to express integrals in source terms, the expression 
of the vector $\vec{\omega} $ in the new system of coordinates is now required. 
We provide the Cartesian components $\omega_x,\omega_y,\omega_z $
of the vector $\vec{\omega}$ as function of $ \theta,\phi,
\Theta $ and $ \Phi $:
\be\label{omegax}
\omega_x= 
\cos \Theta \sin \theta \cos \phi +\sin \Theta \cos \Phi
\cos \theta \cos \phi -\sin \Theta \sin \Phi \sin \phi
\ee
\be\label{omegay}
\omega_y=  
\cos \Theta \sin \theta \sin \phi +\sin \Theta \cos \Phi \cos \theta \sin\phi
+\sin \Theta \sin \Phi \cos \phi
\ee
\be\label{omegaz}
\omega_z=
\cos \Theta \cos \theta - \sin \Theta \cos \Phi \sin \theta.
\ee

The following properties hold
\footnote{In the Cartesian framework, the identities given by Eq.(\ref{prop2}) are quite
trivial: consider the Liouville operator ${\cal L}_{cart}$ in Cartesian coordinate
and Cartesian components:
 $${\cal L}_{cart} = v^i
\frac{\partial F}{\partial x_i} $$
 For $F=p_x$, $F=p_y$ or $F=p_z$ the above identities are fulfilled. 
This obviously holds then for any generic system of
coordinates. Numerically, 
the relations in Eq.(\ref{prop2}),or Eq.(\ref{propf}) can be used to assess the numerical accuracy of our resolution.}:
\be\label{prop1}
\omega_x^2 + \omega_y^2 + \omega_z^2 =1
\ee
\be\label{prop2}
{\cal L}_{sph} \, \omega_x=0, \; \; \; {\cal L}_{sph} \, \omega_y=0 \: \; \;
{\cal L}_{sph} \, \omega_z =0.
\ee

Therefore, if the distribution function $F$ depends only
on $\vec{\omega}$, we have
\be\label{propf}
 {\cal L}_{sph}  \, F(\omega_x, \omega_y, \omega_z)=0. 
\ee

We shall finish this section by noticing that some terms of the
Liouville operator
${\cal L}_{sph}$ given by the Eq.(\ref{Liouvsph}) are singular for  
$r=0$ and  $\theta=0,\pi$. Since the operator is itself regular, 
these terms correspond to coordinate singularities that shall 
cancel each other in the computation. We shall give an example 
of such cancellations in our case. Consider a spherical shell 
in physical space, for which $R_{1} \leq r \leq R_{2}$ and 
$R_{1}>0$. Only singularity issues in  $\theta=0,\pi$ are then to
consider. We first write a polynomial decomposition of the distribution in Cartesian-like coordinates:
\be\label{expansion}
F(x,y,z,\omega_x,\omega_x,\omega_z,t)=\sum_{i,j,k,A,B,C}
C(t)_{ijkABC}x^i y^j z^k \omega_x^A \omega_y^B \omega_z^C.    
\ee

In 6-D spherical coordinates, the singular terms in the Liouville operator given by the Eq.(\ref{Liouvsph}) 
are 
\be\label{sing}
\frac{ \sin \Theta}{\sin \theta} \cos \Phi  \left(
\frac{\partial}{\partial \phi} - \cos \theta
\frac{\partial}{\partial \Phi} \right).
\ee

In the above polynomial decomposition, we encounter two cases:

- For terms associated with coefficients of type $C(t)_{ijk00}$ (no
 dependence on $\omega$),
 a spherical decomposition in $(r, \theta, \phi)$ will lead to $\phi$-dependent terms being factored by $sin(\theta)$.

- For terms containing powers of $\omega_x,\omega_y, \omega_z$, their
expression in 
Eqs.(\ref{omegax},\ref{omegay},\ref{omegaz}) ensures us that compensation 
will occur when the operator in Eq.(\ref{sing}) is applied. 

The spectral representation of the considered fields is able to
  handle directly the specifics of the decomposition (see
  ~\cite{Bona1985, Gran2009} for similar examples). 
\subsection{A simplified 2-dimensional case: The $\Theta=\pi/2$ discontinuity problem} 
 We illustrate the prominent difficulties encountered in the analysis 
of this equation with a problem restricted to a spherically 
symmetric shell ($R_{1} \leq r \leq R_{2}$) and with only coherent 
scattering allowed. The solution $F$ for the distribution function
will then only depend on the three variables $t, r, \Theta$. 
We focus here on analyticity issues and the problem of boundary conditions.
Under the above hypotheses, the transport equation simplifies to:
\begin{eqnarray}\label{Sphersy}
\frac{1}{c} \frac{\partial F}{\partial t}+\cos \Theta 
\frac{\partial F}{\partial r} -\frac{\sin \Theta}{r} \frac{\partial \,
  F}{\partial \Theta} + n_e(r)\left(\sigma_{tot} F(t,r,\Theta) \right. \nonumber \\
 \left. - \int_{0}^\pi 
\hat{\sigma}_{dif} (\cos \Theta \cos \Theta^{'}) \,  F(t,r,\Theta^{'}) \sin \Theta^{'}
\, d \Theta^{'}  \right)=0,
\end{eqnarray}
where $n_e(r)$ is a plasma density, $\sigma_{tot}$ and $\sigma_{dif} $ are respectively the total and
 differential cross section, and integration on the momentum angle $\Phi^{'}$ has already been performed.
 In order to perform a very simple analysis, we now artificially
 split the differential 
operator acting on $F$, so that we retrieve two advection equations. The radial advection part reads: 
\be\label{Sphedif}
\frac{1}{c} \frac{\partial F}{\partial t}+ \cos\Theta \,\frac{\partial F}{\partial  r}=0.
\ee
This is a first order equation, associated to an evolution with
velocity $V= c \cos \Theta $. It propagates from the inner region of the shell to the
outer one if $0 \le \Theta < \pi/2$  ($ \cos \theta > 0 $). 
On the contrary, it propagates from the outer region to the inner one
if $\cos \Theta < 0 $. Consequently, in our geometrical setting, an inner boundary condition at $r=R_1$ has to be imposed for $\Theta \leq \frac{\pi}{2} $
(incoming flux) and an outer condition at $r=R_2$ for $\Theta > \pi/2$ (re-entering flux). 
  
\begin{figure}
\center
\includegraphics[width=0.6\textwidth]{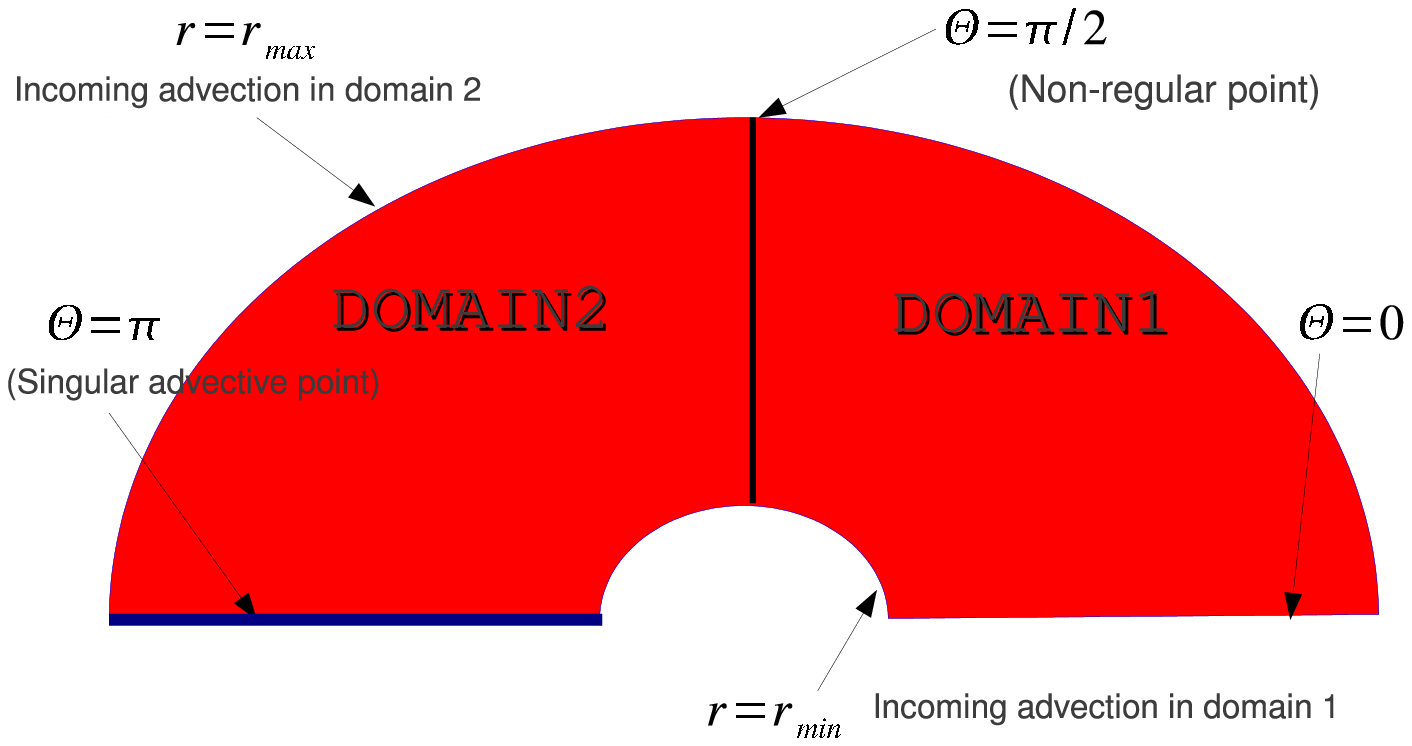} 
\caption{2d representation of the computational domain.}
\label{f:Dom2d}
\end{figure}

If we now consider the second advection term
\be\label{advecthe}
\frac{1}{c} \frac{ \partial F}{\partial t} -\frac{\sin \Theta}{r}
\frac{\partial F}{\partial \Theta}=0,
\ee
the analysis is here simpler: propagation occurs always from
$\Theta= \pi$ to $\Theta=0$ in the computational domain. However, the
vanishing of $\sin \Theta$ at $\Theta=\pi$ shows a degenerate
behaviour at this point: no advection in $\Theta$ occurs, therefore no
boundary treatment is needed. \\ 
Coming back to the full Eq.(\ref{Sphersy}), it is now expected that 
regularity issues in the numerical solution will arise\footnote{We describe a function $F$ as regular if it is of
   class $C^p$ with $p$ large enough to have a fast convergence in the
   spectral expansion.} across the surface $\Theta=\frac{\pi}{2}$, due to different radial
 advective directions on both sides. 
For example, a boundary condition value for $F$ can be freely set to
\be
F(t, R_{1}, \Theta \leq \frac{\pi}{2})=A, A\in {\mathbb{R}},
\ee
whereas the values $F(t, R_{1}, \Theta \geq \frac{\pi}{2})$ are advected 
from the computational domain and therefore uncontrolled.
To overcome the numerical problems associated with this behaviour, we split 
our computational domain (here, a spherical shell) into two angular 
domains $D_{1}(r \in [R_{1},R_{2}],\Theta \in[0, \frac{\pi}{2}] )$ 
 and $ D_{2}(r \in [R_{1},R_{2}],\Theta \in[\frac{\pi}{2}, \pi] )$ (see
Fig.~\ref{f:Dom2d}). 
To ensure particle number conservation across the two domains,  
we must enforce continuity of the flux on $\Theta=\frac {\pi}{2}$. 
This provides us with an incoming boundary condition in $\Theta$ to impose for the solution $F$ in $D_{1}$.

\section{Numerical tests}
\label{sect:numtest}

\subsection{Overview of the computational setting and approach}
We present below specific tests related to the spectral resolution of
the (homogeneous or not) transport equation. As outlined above, our
computational grid covers a physical shell $R_{1}\leq r \leq R_{2}$
(see section~\ref{sect:rotns} for the treatment of the singularity at the
center) split into two domains $D_{1}[0 \leq \Theta \leq
\frac{\pi}{2}]$ and $D_{2}[\frac{\pi}{2} \leq \Theta \leq \pi]$. A
typical value for our domain size is $\frac{R_{2}}{R_{1}}=5$. Unless
otherwise stated, spectral decompositions are performed using a
Chebychev representation on the $r$,$\Theta$ and $\gamma$ direction,
whereas a Fourier decomposition is performed for the remaining angular
dependencies. The spectral decomposition of a scalar field is then
very much similar to the one described in \citep{Bona1990}, however
performed in six dimensions instead of the usual three. For general
information on numerical use of spectral methods as intended here, we
direct the reader to the recent review of \citep{Gran2009}.

 If we denote by $T_{i}$ the ${i^{th}}$ order polynomial in the classical Chebyshev basis and by $Y_{\ell m}(\theta, \varphi)$ the 3D spherical harmonics component of order $(\ell,m)$, a decomposition of the 6D time-dependent distribution function is given by:
\begin{eqnarray}
F(r,\theta,\varphi,\gamma, \Theta, \Phi,t)= \nonumber \\
\sum_{i,\ell,m,A,B,C}C_{i\ell m ABC}(t)T_{i}(r)Y_{\ell m}(\theta,\varphi)T_{A}(\gamma)T_{B}(cos(\Theta))&cos(C\Phi), C \textrm{even}, \nonumber \\
&sin(C\Phi), C \textrm{odd}.\nonumber
\end{eqnarray}

where we manipulate the set of coefficients $C_{i\ell mABC}(t)$ as the representation of $f(r, \theta, \phi, \nu, \Theta, \Phi,t)$ at any time. The representation above assumes a symmetry with respect to the $(\omega_{y},\omega_{z})$ plane to obtain this particular dependence in $\Phi$. Otherwise, all terms of the Fourier decomposition have to be considered.
  All numerical operations are then performed in the coefficient space, and using the product base described in the above expansion. Imposition of boundaries is performed using a Tau approach \citep{Gott1977}. In particular, differential functions composing the Liouville operator are expressed as matrices acting on the coefficient vectors $C_{i\ell m ABC}$. A semi-implicit resolution in the Appendix B also uses Tau like methods for operator inversion, handling numerically vectors of spectral coefficients $C_{i \ell m ABC}$. 

 In this section, the chosen explicit time marching scheme is a classical second order Adams-Bashforth one, minimizing dissipation. Again, only spectral coefficients are updated. 

The chosen computational domain is the shell set of domains described in the previous section ; Tau-matching is performed at the innermost and outermost sphere, as well as at the $\Theta=\pi/2$ interface. In the diffusion transport problem of section 6, a central sphere-like domain is added to the setting, in which the representation of functions is the same as in the rest, and for which numerical solutions have of course to be matched through the outer interface (see again Section 6).

\subsection{Time evolution of a uniform distribution}
  We assume our domain to be filled by a uniform plasma of constant density
  $n=n_{0}$, which at first is interacting with our radiating particles only through coherent
  scattering. Absorption and emission are 
disabled (which ensures particle number conservation during the computation)
  and we start with the artificial 
initial condition for the distribution function:
\begin{eqnarray}
F(\theta, \phi, \Theta, \Phi,t=0)=\left( 1 + 2\omega_{x} + \omega_{y}
+ \frac{1}{2} \omega_{z}\right)^{4}, 
\; \; \; 0 \le \Theta \le \frac{\pi}{2} \nonumber \\
F(\theta, \phi, \Theta, \Phi,t=0) =0,\; \; \; \frac{\pi}{2} \le \Theta
\le \pi.
\end{eqnarray}
Taking advantage of the properties of the Liouville operator described
in Eq.~(\ref{prop2}), we know that at any time of the computation,
$\mathcal{L}_{sph}F=0$. Monitoring the numerical validity of this
property is another way to assess accuracy of our approach.
\begin{figure}
\center
\includegraphics[width=0.45\textwidth]{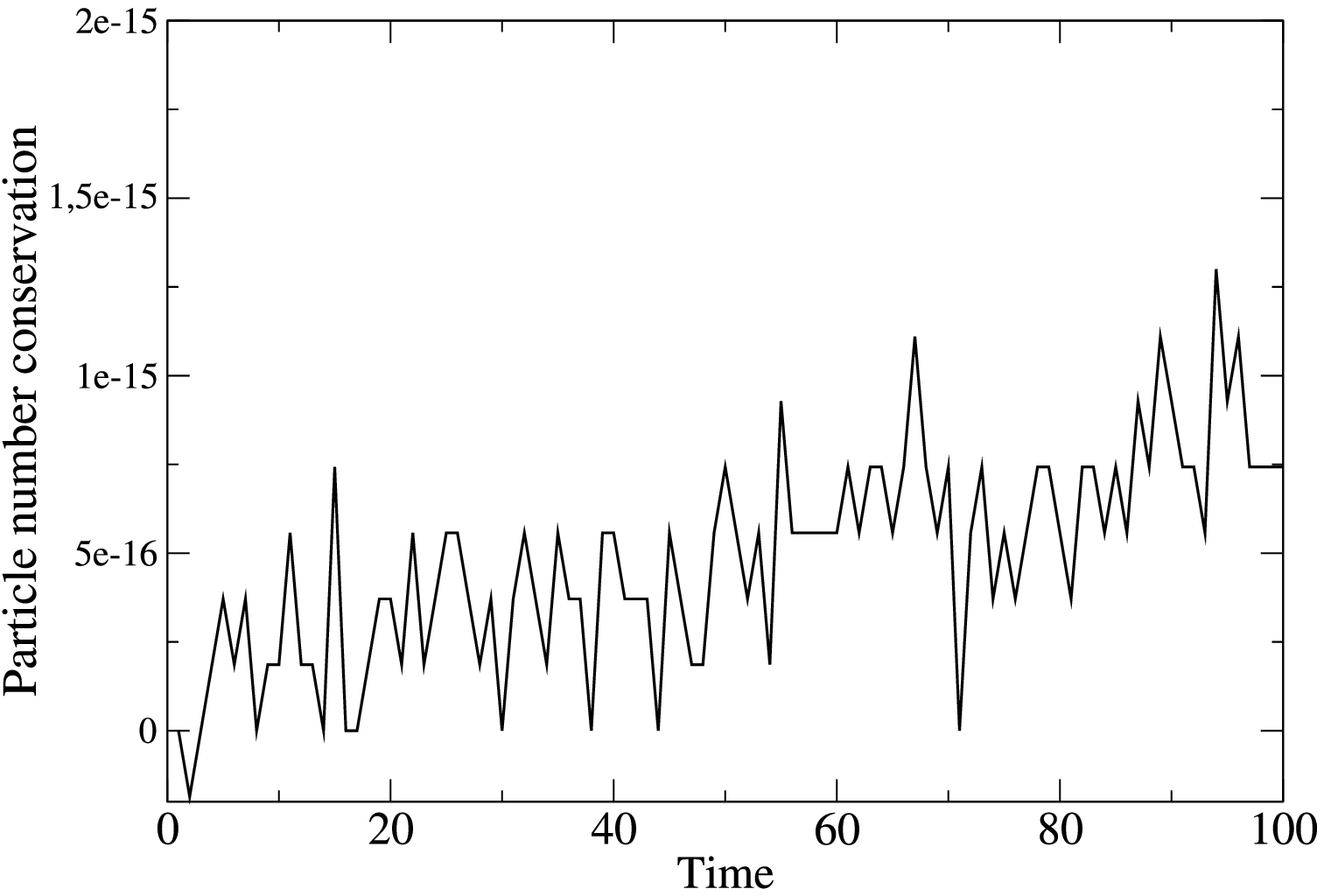} 
\caption{Relative particle number conservation for the coherent time evolution of a uniform distribution}
\label{f:istro4}
\end{figure}
\begin{figure}
\center
\includegraphics[width=0.45\textwidth]{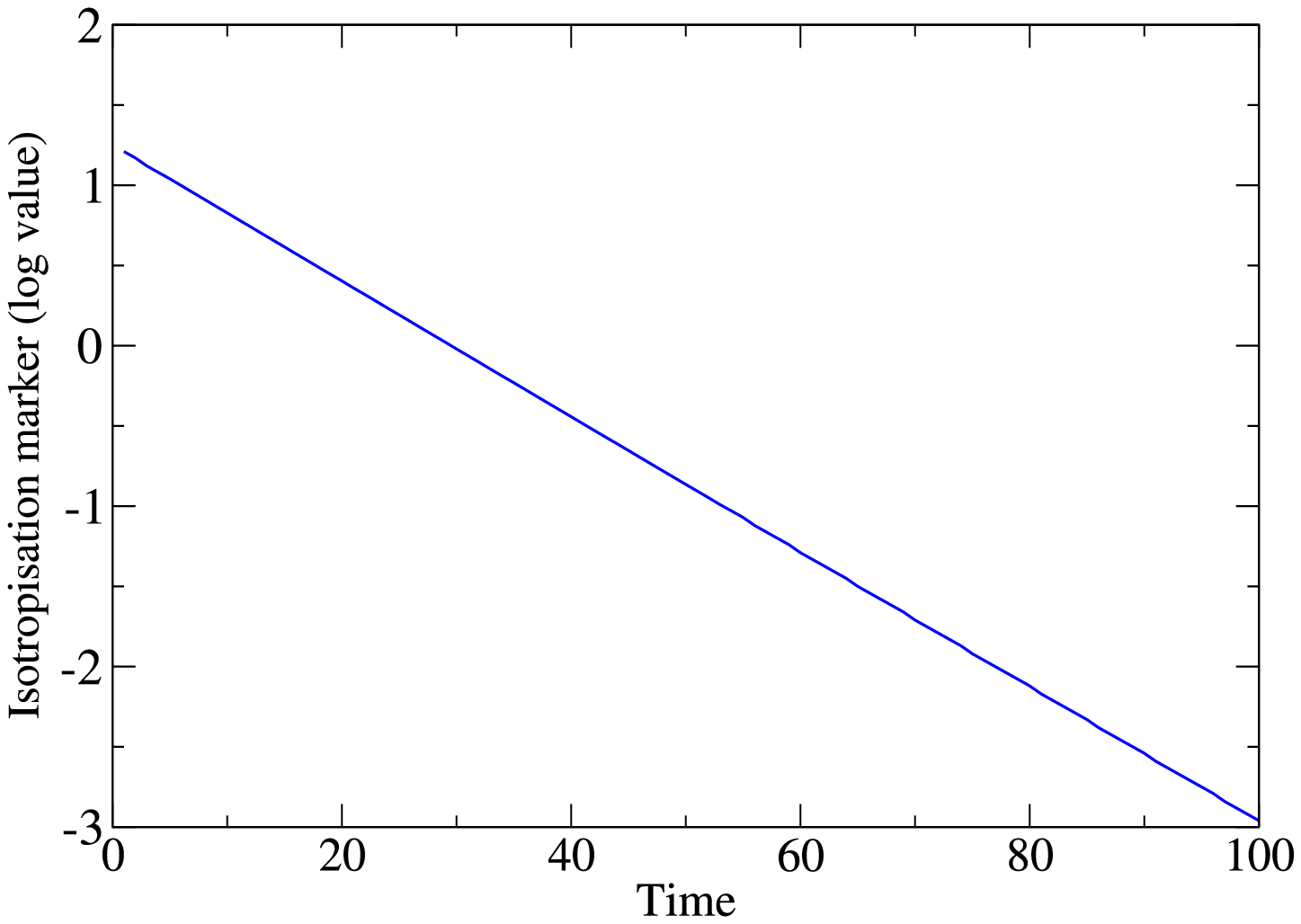} 
\caption{$L^{0}$ norm of the non-isotropic terms in the spectral
  decomposition of the distribution function: this is computed as the  ratio 
between the coefficient of the constant (isotropic) term in the  decomposition, 
rescaled to the sum of all other coefficients. This serves as a  reliable 
marker for the isotropisation process in the evolution.}
\label{f:istro6}
\end{figure}

Fig~\ref{f:istro4} presents particle conservation for this setting
over time. The slow drift we encounter only occurs at the level of
computer roundoff. We consistently obtain a relative error in particle
number count smaller than $5.10^{-15}$ in double precision, on
timescales much larger than the dynamical timescale of the
simulation. The isotropisation process of the distribution function
due to coherent scattering is also displayed on
Fig~\ref{f:istro6}. For those results, the number of points used is
$(N_{r},N_{\theta},N_{\phi},N_{\Theta},N_{\Phi})=(33,17,16,25,16)$. 
A resolution time step takes about 20 seconds in CPU time. 

 Using the same initial spatial profile for the distribution, we now
 allow for energy dependence and 
non-coherent scattering by implementing the energy-dependent source
 terms set in Eq.~(\ref{Transpfp}). 
The initial energy distribution is set to be a black body one, at a
 temperature half the one of 
the plasma ($k T/m_ec^2=.01$ in our units). Conservation of the number
 of photons ensures that $F$ will 
approach a Bose distribution (see Eq.~(\ref{Bose})) over time. Using
 $N_{\gamma}=33$ points in the energy 
dimension, a computational time step takes about 33s. 

\begin{figure}[h]
\begin{tabular}{ccc}
\noindent \includegraphics[width=0.45\textwidth]{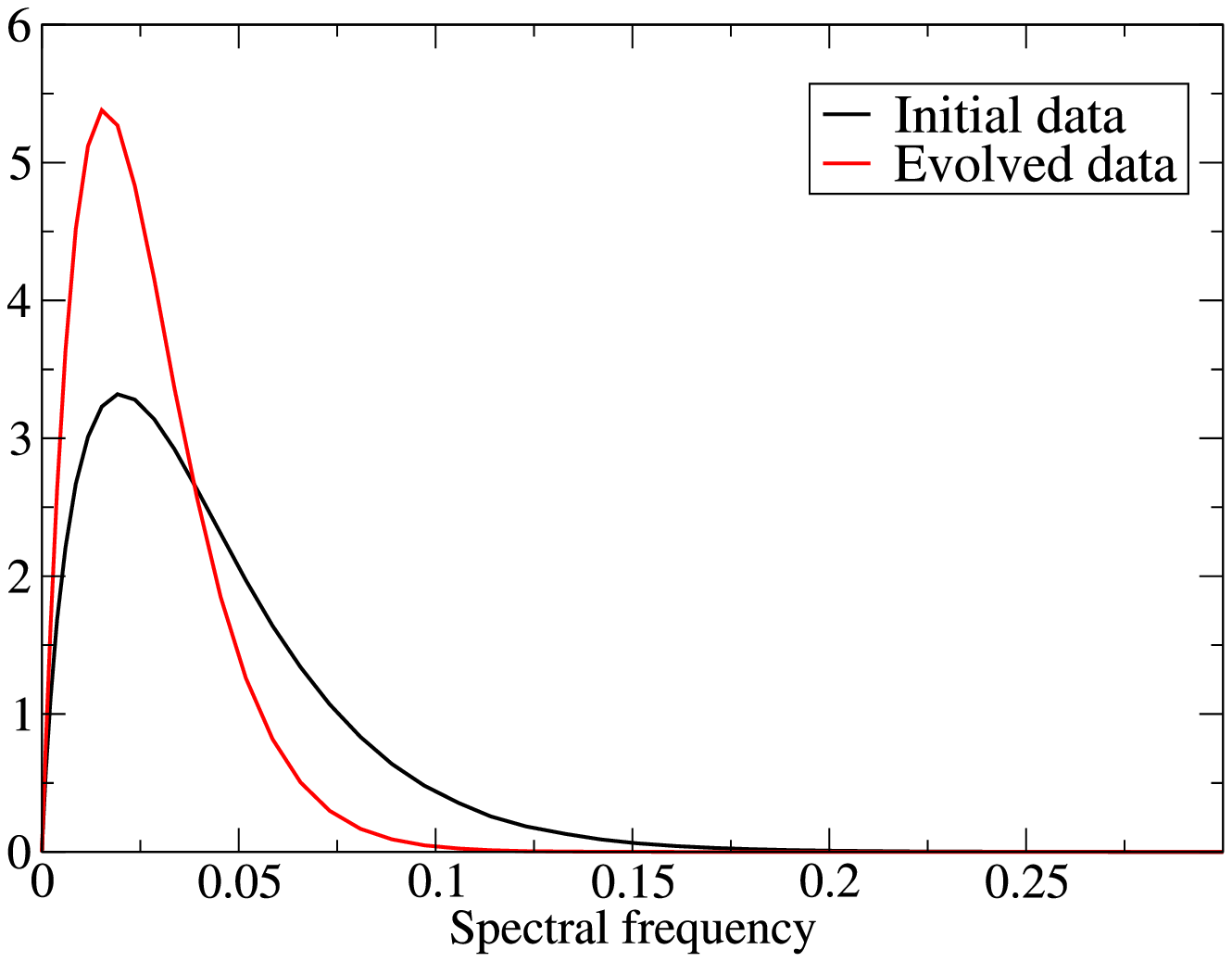}
\\
\includegraphics[width=0.45\textwidth]{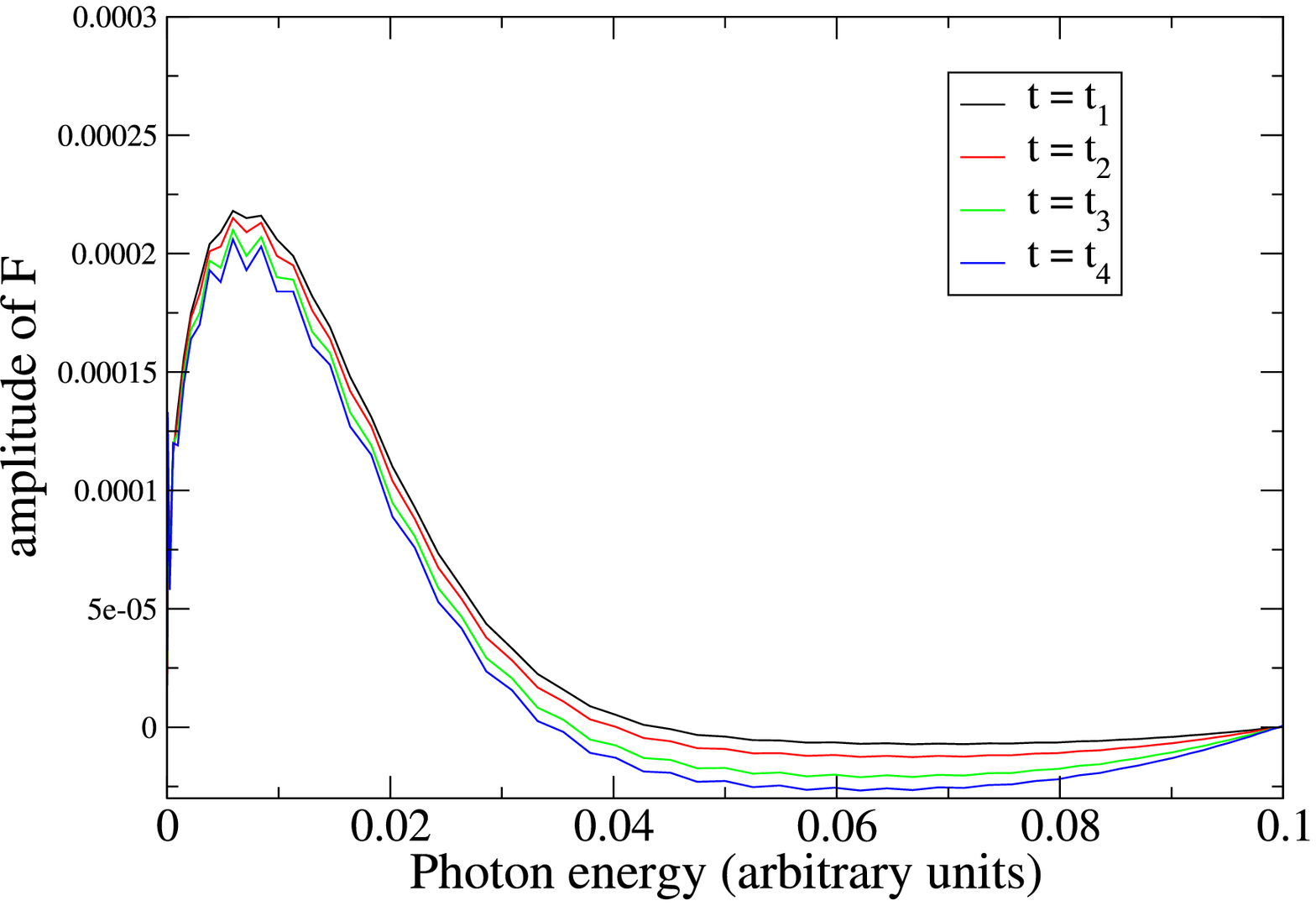}
\\
\includegraphics[width=0.45\textwidth]{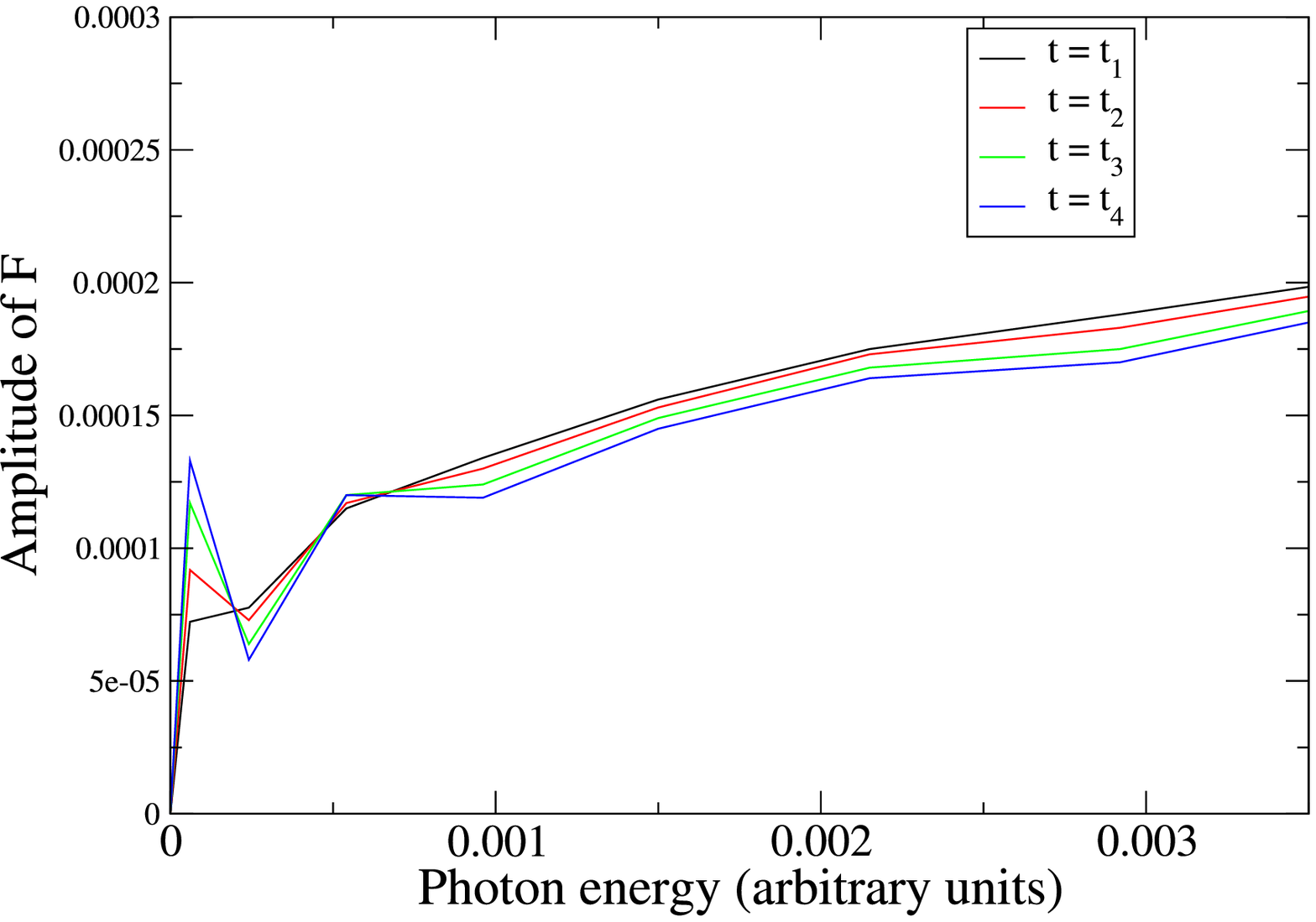}
\end{tabular} 
\caption{Evolution of the energy distribution of $F$, showing the
  transition to a Bose 
distribution and a low-energy condensation. The middle and lower panel
  display this energy distribution 
at four arbitrary consecutive times. No smoothing of the data has been performed.}
\label{f:fokker3}
\end{figure}

In Fig.~\ref{f:fokker3} we can observe the transition made from the
initial energy distribution to the final 
one, and appreciate the possible observation of a low-energy
condensation that is accessible even with a 
very limited number of points. It is obvious that a specific treatment
of the low energy regime
 (by allocating a specific spectral decomposition domain to this
region, and increasing the degree of 
spectral decomposition) would be necessary to study such an effect
quantitatively ; however, the goal of this work 
is only to convince oneself that such study is, indeed, possible with limited computational resources.
\subsection{5d coherent transport in a shell} 
 
 We consider a spherical shell enclosing black body radiating
 particles
through its outer surface $R=R_{2}$. In the computed shell domain
 resides a plasma with the following 
arbitrary density:
\begin{equation}
n(r,\theta,\phi)=n_0 [1+0.1 (r \sin \theta \cos \phi \cos \theta)]
\left [1-(1-r/R_1)/(1-r/R_{2}) \right]^8. 
\end{equation}

 This plasma triggers coherent scattering, but again emission and
 absorption processes are disabled 
for simplicity: we only want to monitor the behaviour of a transport
 process. The shell is initially 
free of any radiating particles, and the central object emits
 continuously a particle flux following a Lambert law; this leads to the inner boundary condition for $F$:
\be
F(R_{1},\theta,\phi,\Theta,\Phi)=F_0 \cos \Theta.
\ee

This problem will be treated spectrally, except for the radial
 direction $r$ where we use a simple first
order finite-difference scheme. The reason behind it is a better
 treatment of the discontinuity and a reduction of the overshooting in this direction that inevitably appears.
Grid point numbers are
 $(N_{r},N_{\theta},N_{\phi},N_{\Theta},N_{\Phi})=(129,17,16,25,16)$, and a time step is around 216s wall clock time, again on a single core.
\begin{figure*}
\begin{tabular}{cc}
\noindent \includegraphics[width=0.5\textwidth]{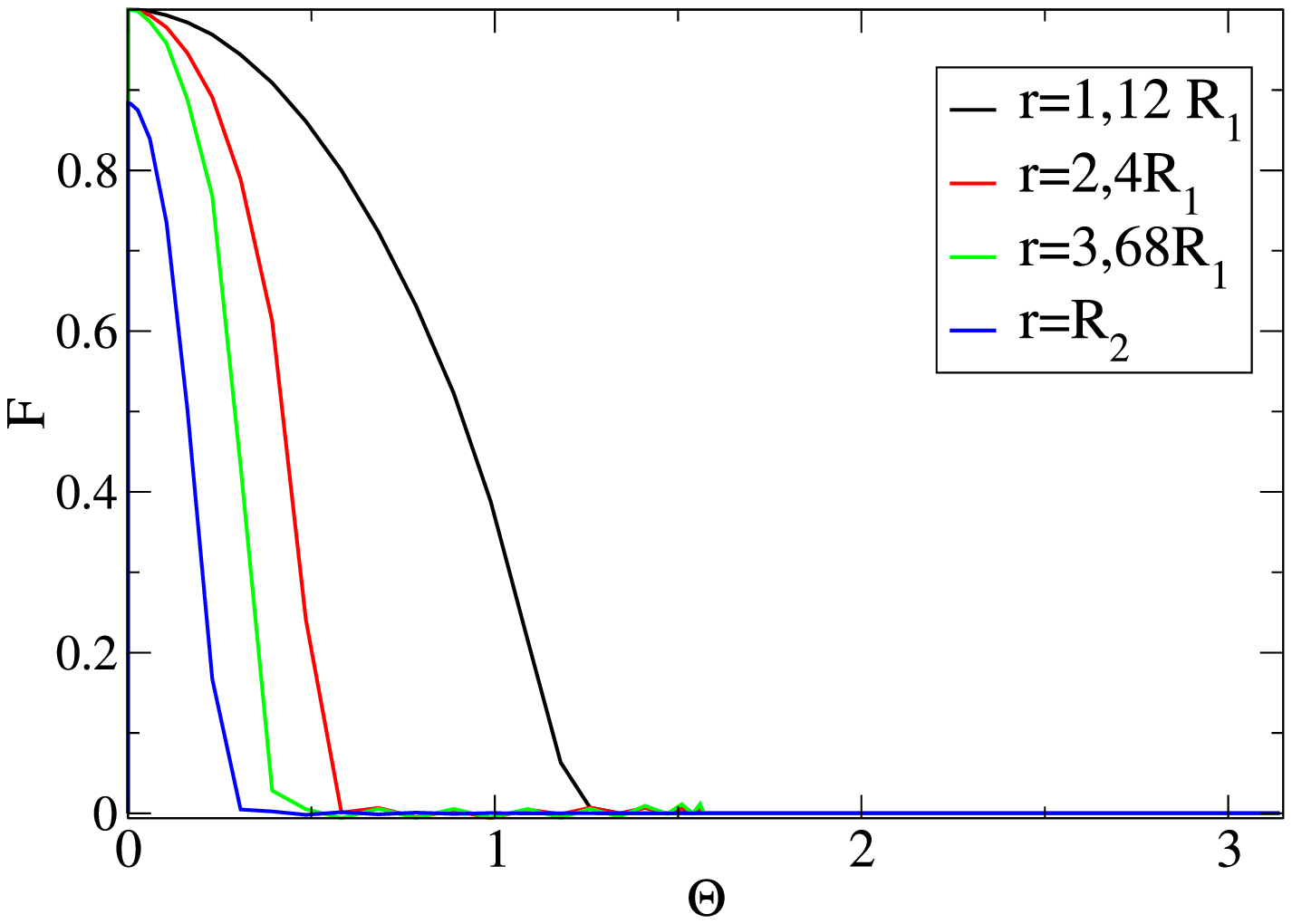}
&
\includegraphics[width=0.5\textwidth]{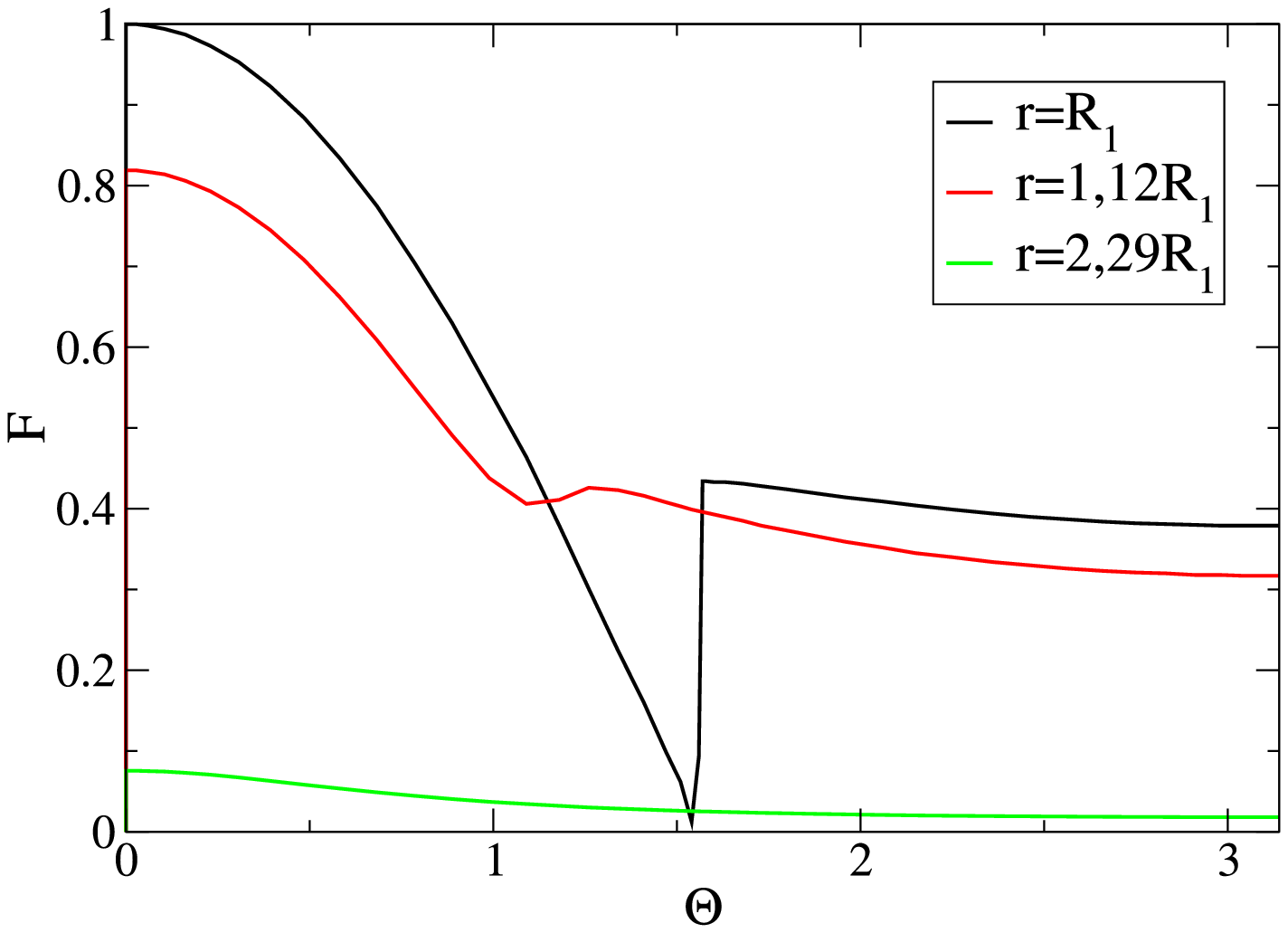}
\end{tabular}
 \begin{tabular}{cc}
\noindent \includegraphics[width=0.5\textwidth]{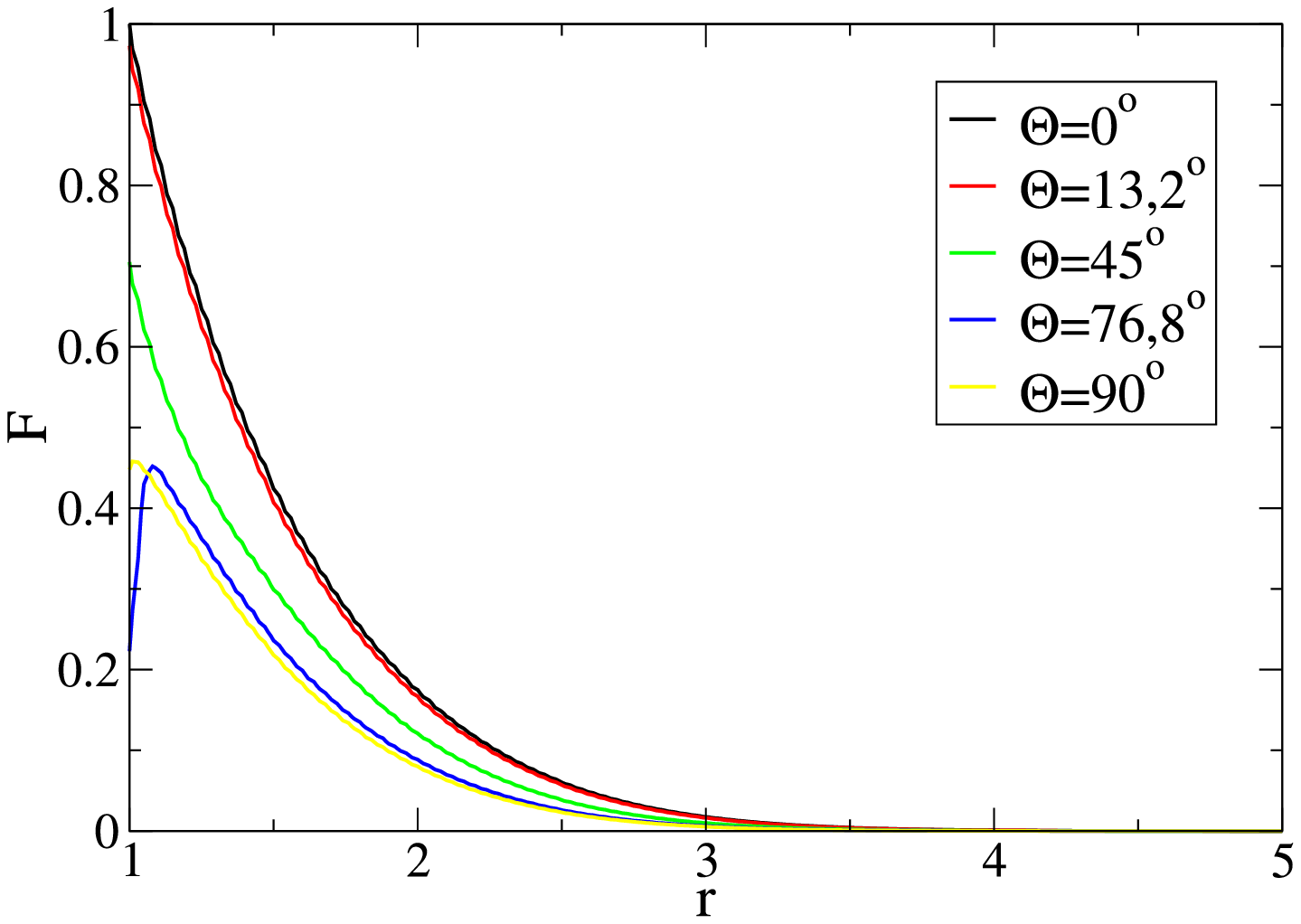}
&
\includegraphics[width=0.5\textwidth]{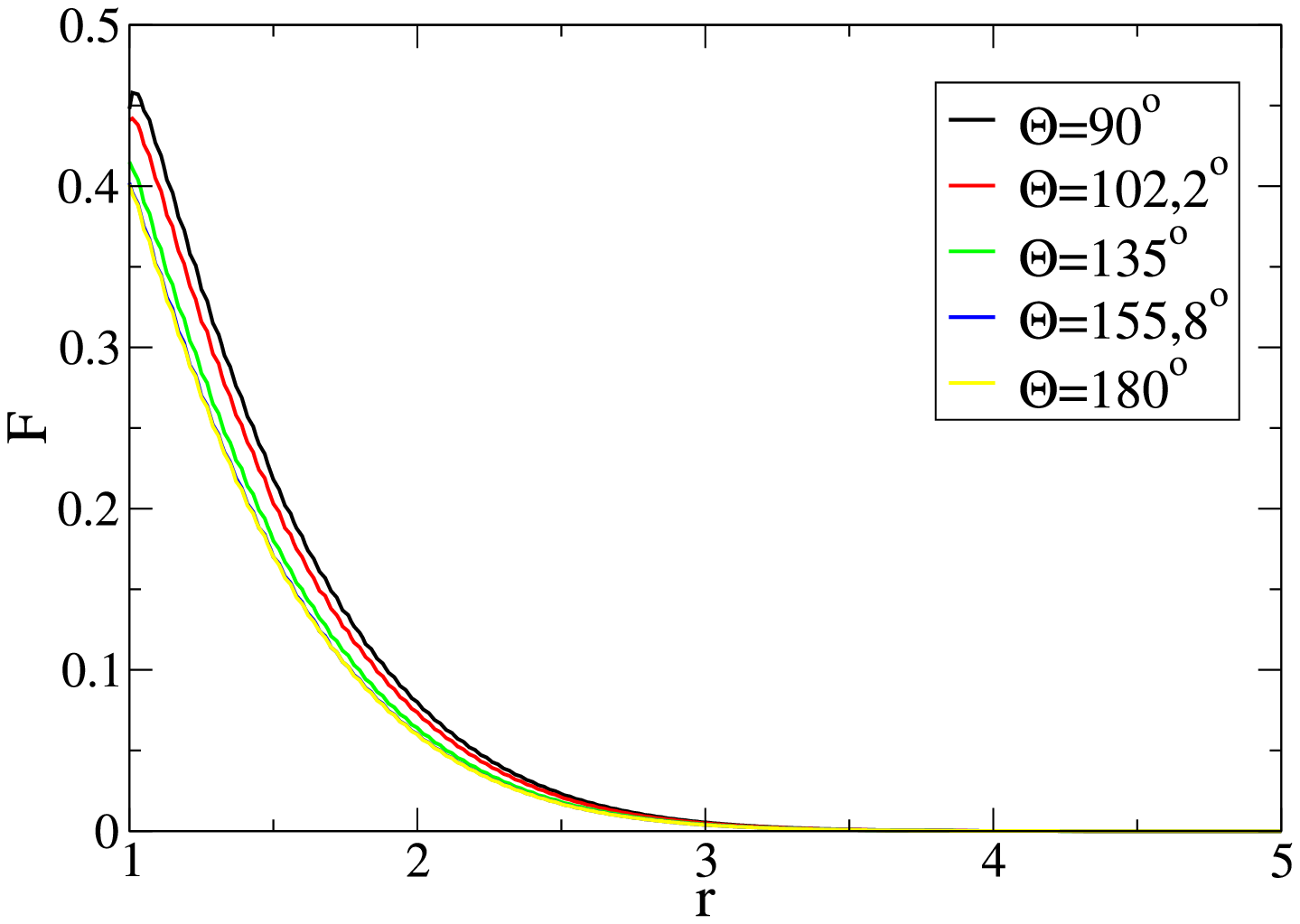}
\end{tabular}
\caption{Profiles of the distribution function $F$ along the
  directions $r$ and $\Theta$ at time $t=4,2$ in our units. The upper left panel
  represents angular profiles assuming zero optical depth for coherent scattering, whereas the upper right one corresponds to an optical 
depth of 5 in our units. The two lower panels are radial profiles at 
different angles for the same time, and for an optical depth of $5$. 
The boundary values correspond to the thermal emission of a Lambertian  object 
(a perfect black body here), and a non-re entrant condition for the outer radius.}
\label{f:ADV}
\end{figure*}
 Fig~\ref{f:ADV} presents distribution function profiles at different
 time steps in the case of 
an optically thin regime (optical depth with respect to the coherent
 scattering is set to zero), 
or an optically thick one (in our arbitrary units, the optical depth
 is set to 5). 
We are able to represent without any problem the beaming effect occurring
 during the 5d transport, 
which is also coupled to a large attenuation in the second case 
(the loss of luminosity by 5 orders of magnitudes on a short distance
 is not altering the code precision, 
as can also be seen by a check on the particle number conservation). 
It is obvious that in the transparent case, an excessive beaming will
 eventually lead to resolution 
issues in the angular directions ; this issue can be cured again (at
 least locally) by treating a low 
$\Theta$ region separately (domain decomposition for the spectral
 treatment) and assuming a 
better resolution at small angles. It is obvious that an open free streaming 
region cannot be handled by our approach in a clean way : one would
 have then to resort to less sophisticated descriptions of the particle flux, and match to the exact solver.
We observe a clear discontinuity of $F$ at the edge $r=R_{1}$ 
of the domain, assuming a non-zero optical depth; it is a consequence 
of our rather abrupt assertion for the radiation source to be a pure 
Lambertian object. A more sophisticated approach for the source would 
allow to get rid of such a feature, although this computation proves 
that the code behaves well even in ill-posed settings.

\subsection{Stability and convergence conditions}

We end this section by expliciting the stability conditions constraining the model evolution. In general, the maximal allowed value for the time increment $\Delta t$ is determined by the most stringent Courant condition in each dimension. In the 5-D hybrid advection code above, the stability condition gives an order of the timestep limit as the minimum of the following values:
\be
 \Delta t_r =
\frac{(R_2-R_1)}{N_r^{2}},  \Delta t_{\theta} = \frac{1}{N_{\theta}} ,
 \Delta t_{\phi} = \frac{1}{N_{\phi}},  \Delta t_{\Theta} = \frac{1}{N_{\Theta}^2}
, \Delta t_{\Phi} = \frac{1}{N_{\Phi}},
\ee
with notations introduced above. The most severe limitations are given in the $r$ and the $\Theta$ dimensions. In the example showed in Fig.\ref{f:ADV}, the timestep limit is $\Delta t=5 \times 10^{-3}$ (2000 timesteps) for $N_r=129$, $N_\Theta=17$.   

\section{An example: neutrino transfer in a rotating neutron star}
\label{sect:rotns}

The neutrino transfer in a hot rotating neutron star is a first step
towards solving the problem of a cooling neutron star (NS).  We
consider a slowly rotating steady state neutron star with a given
neutrino distribution and we apply all the machinery described above
in computing the neutrino flux as function of time. We consider this
academic example simple enough but containing difficulties that are
present in a wide class of problems, like the evolution of a
proto-neutron star (PNS) for which hydrodynamics and neutrino transfer
are coupled. Once again, the emission and absorption coefficients are
omitted because their presence does not add numerical difficulties,
and their absence allows us to test performances of the code like
neutrino conservation, continuity of the solution. In fact emission
and absorption terms can hide defects of the solution.
\\
\\
\\
In what follows, we consider a slowly rotating NS (The surface of the
NS is only weakly deformed by the rotation), on an axisymmetric
equilibrium configuration determined by an arbitrary equation of
state.

Because of the symmetry, the transfer problem reduces to a 5-D
problem: the variables are $r, \; \theta, \; \Theta, \; \Phi $, the
energy of the neutrinos $E$ plus the time $t$.

In the collision operator, only the nucleon scattering is taken into
account, and the approximation of coherent scattering is used.

Because of the slow rotation, the plasma can be considered at
rest. Note that this approximation also holds for a PNS cooling and
shrinking gently.  For a fast rotating neutron star, described in a
next section, we shall give an hint on how to treat the problem.

As already said, this example is used to show different difficulties
that are present in solving the above problem and how to overcome
them. For simplicity and without any loss of generality we have chosen
an analytic profile of mass density

\be
\label{dens} n(r,\theta)= n_0
\left[ 1-\left( \frac{r}{R_*} \right)^2+\frac{1}{2} \, \left(\frac{2
      \pi}{P} \frac{r}{c} \sin \theta \right)^2 \,
\right]^{1/(\gamma-1)}~,
\ee 

where $n_0$ is the central density, $R_*$ the radius of the star, $c$
the light velocity, $P$ the rotation period and $\gamma$ mimics the
polytropic index ($ 4/3 \le \gamma \le 2$). For $n_0=4 \times 10^{14}
\ \textrm{g.cm}^{-3}$, $R_*=15 \ \textrm{km}$, $\gamma=5/3$ and $P \ge
3 \ \textrm{ms}$, the mass of the star is $ 1.4 \ \textrm{M}_{\odot}
$.  The above analytic expression of the matter density has the
advantage to be flexible and to mimic different equations of state.
For realistic applications, the profile density must be computed by
solving the G.R. equations for a steady state configuration. In this
case, the value of the mass density $n(r,\theta) $ is given in the
sampling points of the variables $r$ and $\theta$.  The numerical grid
defined in a such a way will be called the {\it
  master grid}. 

\vspace{1em}
 
The main difficulties addressed in our example are:

\begin{enumerate}
\item How to handle the singularity at the center.
\item How to treat the large range of opacities, especially the
  strong dependence on the neutrino energy $E$.  Recall that the
  neutrino cross section behaves as $\sim E^2$
\end{enumerate}

We have not yet solved the $r=0$ singularity problem in the above
formalism. Moreover, the opacity $\tau=n \, \sigma_t$ ($\sigma_t$ being
the total scattering cross section) close to the center of the star can be so
large that it would require an excessively small time step.

In order to bypass this difficulty we propose to cut the $r$ domain in
two regions : a first one running from $0$ to $R_{in}$ ($0 \le r \le
R_{in}$) and a second one running from $R_{in}$ to $R_*$ ( $R_1 \le r
\le R_{im}$) The value of $R_{in}(E) $ is determined in a such a way
that for a given energy $E$ the opacity $\tau$ in the domain $0 \le r
\le R_{in} $ is larger than a critical value $\tau_c$ for which the
diffusion approximation holds. Numerical experiments have shown that in
the above example, $\tau$ must be

\be
\label{taud}
\tau_c \ge 2 \ \textrm{km}^{-1}~,
\ee

in order that the diffusion approximation holds and and

\be
\label{taue}
\tau \le 10 \ \textrm{km}^{-1}~,
\ee 

to have an acceptable time step to solve the exact transfer equation.

The way to proceed is the following : solve the diffusion equation
(See App.~\ref{app:diffeq})

\begin{eqnarray}
\label{diffusion}
\lefteqn{\frac{\partial F_0(r,\theta,E,t)}{\partial t} - \tau(r,\theta,E,t) 
\Delta F_0(r,\theta,E,t)}\nonumber\\
&& -\nabla_j F_0(r,\theta,E,t) \nabla^j
\tau(r,\theta)=S(r,\theta,E,t)~,
\end{eqnarray}

in the domain $0 \le r \le R_{in}$. Note that, as it was already said,
the absorbtion and emission terms in $S$ is put to $0$ in the present
example. Then solve the exact transfer equation in the domain $R_1 \le
R_*$ and then match the two solutions (Here $S(r,\theta,E,t) $ are the
source terms generated by absorption and emission and $\Delta$ is the
Laplacian in spherical coordinates.

\be
\label{Laplacian}
\Delta= \frac{\partial^2}{\partial r^2}+\frac{2}{r}
\frac{\partial}{\partial r} + \frac{1}{r^2} \left(\frac{\partial^2}
  {\partial \, \theta^2} + \frac{\cos \theta}{\sin \theta}
  \frac{\partial}{\partial \theta} \right).
\ee

The diffusion equation is solved with a
  semi-implicit scheme \citep{Gott1977}. After
  expansion in Legendre polynomials $P^0_l(\theta) $ a second order
  time scheme can be written with obvious notations at the time
  $t^{j+1} $ for a given energy $E$

  \begin{eqnarray}    
  \label{difimp}
  \lefteqn{F^{j+1}_{0 \,l}(r,E) -\tau_{max} \Delta_l F^{j+1}_{0 \, l} \frac{dt}{2}
  =F^{j}_{0 \,l} -\frac{dt}{2} \tau_{max} \Delta_l F^{j}_{0 \,l}
  }\nonumber\\
  &&+\left[ \left( \tau(r,\theta \, E)-\tau_{max}\right) \Delta F_0
    -\nabla_j F_0 \nabla ^j \tau + S \right]^{j+1/2}_l dt~,
  \end{eqnarray}

  where

  \be
  \label{Laplacl}
  \Delta_l= \frac{ d^2}{dr^2} + \frac{2}{r}
  \frac{d}{dr}-\frac{1}{r^2} \, l \,(l+1)~,
  \ee

  with $\tau_{max}$ the maximum value of $\tau$ in the domain and $dt
  $ the time step. The terms at time $j+1/2$ are obtained by
  extrapolation using the terms at the time $j-1 $ and $j$ .  The
  singularity at $r=0$ is handled by choosing an expansion on a
  polynomial basis that has good analytical properties at $r=0$. The
  scheme is unconditionally stable (\cite{Bona1990}).  Moreover,
  in Eq.(\ref{difimp}) the matrix of the operator at the L.H.S can be
  reduced to a penta-diagonal matrix.


\subsection{The mono energetic case}

In this section the conservative formulation of the transport equation
will be used and the dependence of $F$ on $E$
will be omited . \\
\\

We take an averaged cross section $\sigma^D$

\be
\label{cross}
\sigma^D= \sigma_{\nu n}^D+\sigma_{\nu p}^D,
\ee 

from \cite{brue_85} : for neutrino neutron scattering

\be
\label{crossnun}
\sigma_{\nu N}^D =A_{\nu \, N} \left(1-
  \frac{1}{3} \vec{\omega}\cdot \vec{\omega}^{'} \right), \; \; \; \;
\sigma_{\nu \, N}^T =4 \, \pi A_{\nu \, N} ,
\ee

and for protons

\be
\label{crossnup}
\sigma_{\nu P}^D =A_{\nu \, P} 
\left(1-\frac{1}{10} \vec{\omega}\cdot \vec{\omega}^{'} \right), \; \;
\; \; \sigma_{\nu P}^T=4 \, \pi A_{\nu \,P} .
\ee

Let us consider some mono energetic neutrinos with $E = 3 \
\textrm{MeV}$. We take $R_{in}=11.4 \ \textrm{km}$. We form two
domains, $0 \le r \le R_{in}$ and $R_{in} \le r \le R_* $. In the
first domain the optical depth is $\tau \ge 5 $. In this domain we
shall solve the diffusion equation by using spectral methods. The grid
in the first domain has a Chebyschev decomposition for $r$ sampling
points (the $\theta$ sampling points are unchanged). In the second
domain a uniform grid is defined. (We use the hybrid version of the
transport equation, i.e. the $r$ dependence of the variables are
treated with a finite difference scheme). We consider a $1.41 \
\textrm{M}_{\odot} $ neutron star with a rotation frequency of $637 \
\textrm{Hz}$ (corresponding to $\Omega=4000 \ \textrm{rad.s}^{-1} $),
and a ploytropic index $\gamma=5/3 $.  The matter density is given by
Eq.(\ref{dens}).

We define the matter density function $n(r,\theta) $ on the two
grids\footnote{In our case, the function $n(r,\theta)$ is analytic,
  but in a general case, one would have to perform an
  interpolation.}. We shall introduce the obvious notations
$n_1(r,\theta)$, $n_2(r,\theta)$ and $F_1$, $F_2$ defining
quantities in the domains $(1)$ and $(2)$. \\

\subsection{Matching}

Matching of the two solutions $F_1$ and $F_2$ at $r=R_{in}$ cannot be
exact. In fact the solution $F_1$ obtained with the diffusion
approximation contains only two moments $F^0$ and $F^1$

\be
\label{Moments}
F_1(r,\theta,\Theta,\Phi,t) = F^0_1 (r,\theta,t)+ 3
\vec{F}^1_1(r,\theta,t) \cdot \vec{\omega}~,
\ee

where (see App.~\ref{app:diffeq})

\be
\label{Fick}
\vec{F}^1_1=-\frac{1}{3\tau}\vec{\nabla} F^0_1~.
\ee

On the contrary, the exact solution $F_2$ contains a large number of
moments, consequently the matching cannot be exact.

To overcome this difficulty we perform an averaged matching that
conserves the number of neutrinos and we impose an averaged continuity
of $F_1$ and $F_2$\footnote{In the same spirit as the Marshak approximation
for imposing boundary conditions.}.

Before we explain the way to proceed, we have to recall that the
solution of the second order diffusion equation admits two homogeneous
solutions $H_1(r,\theta,t) $ and $H_2(r,\theta,t)$. One of the
homogeneous solution is used to handle the coordinate singularity at
$r=0$ (\footnote{ For more details see \cite{Bona1990}.}), the second
one is used to satisfy the boundary conditions at $r=R_{in} $ for each
value of $\theta$ and at each time $t_j$.

As it was already stated, boundary conditions at $r=R_{in} $ can be
imposed on the solution of the full transport equation
$F_2(r,\theta,\Theta,\Phi,t)$ only in the case $0 \le \Theta < \pi/2$
(see section~\ref{sect:mathframework}). At each time step $t_j$ we impose
the following boundary conditions (B.C.) for $F_2$

\be
\label{F2bc}
F_2(R_{in},\theta,\Theta,\Phi,t_j)=F_2(R_{in},\theta,\pi
/2,\Phi,t_j)+\beta \cos \theta,
\ee

where $\beta$ is determined together with the boundary conditions of
$F_1$, so that

\begin{eqnarray}
\label{F1bc}
\lefteqn{\int_0^{2 \pi} d\Phi
\int_0^{\pi} F_1(R_{in},\Theta, \Phi) \, d \Theta} \nonumber\\
&&=\int_{0}^{2\pi} d \Phi \left[ \int_{0}^{\pi/2}
  (F_2(R_{in},\theta,\Theta, \Phi)+\beta \cos \Theta ) d \Theta \right.\nonumber\\
&&\left. +\int_{\pi/2}^{\pi} F_2(R_{in},\theta,\Theta,\Phi) \, d \Theta
\right]~,
\end{eqnarray}

and the flux conservation

\begin{eqnarray}
\label{Fluconser}
\lefteqn{\int_0^{2
  \pi} d\Phi \int_0^{\pi} F_1(R_{in},\theta,\Theta,\Phi) \cos \Theta
\, d \Theta =} \nonumber\\ 
&&\int_0^{2 \pi} d \Phi \left[
  \int_0^{\pi/2} (F_2(R_{in},\theta,\Theta,\Phi)+\beta \cos \Theta )
  \cos \Theta \, d \Theta \right. \nonumber\\
&&\left. +\int_{\pi/2}^{\pi} F_2(R_{in},\theta,\Theta,
  \Phi) \cos \Theta \, d \Theta \, \right].
\end{eqnarray}

By taking into account the Eqs.(\ref{Fick}) and (\ref{Moments}), the
system of equations Eqs.(\ref{F1bc}) and (\ref{Fluconser}) reads
(matching the distribution function)

\begin{eqnarray}
\label{BCsys1}
\lefteqn{4 \pi
F^0_1(R_{in},\theta)=\int_0^{\pi/2} F_2(R_{in},\theta,\pi/2,\Phi) d
\Phi} \nonumber\\ 
&&+ \int_0^{2\pi} d \Phi \int_{\pi/2}^\pi F_2(R_{in},\theta,
\Theta,\Phi) \,d \Theta +\pi \beta,
\end{eqnarray}

and as for the flux conservation

\begin{eqnarray}  
\label{Bcsys2}
\lefteqn{-\frac{4 \pi}{3 \tau} \frac{ \partial
  F_0(R_{in},\theta)}{ \partial r} =\int_0^{2 \pi} d \Phi
_,\left[F_2(R_{in} , \theta, \pi/2, \Phi) \right.} \nonumber\\ 
&&\left. + \int_{\pi/2}^{\pi}
  F_2(R_{in} ,\theta,\Theta,\Phi) \, d \Theta \right] + \frac{2}{3}
\pi \beta.
\end{eqnarray}

As it was already stated, the unknown of the system are the boundary
condition $F_1((R_{in},\theta,t_j) $ and the coefficient $\beta$. Note
that once $F_1(R_{in},\theta,t_j)$ is given, its derivative with
respect to $r$ is known.

\subsection{The multi-energy case}
\label{sec:multienerg}

To treat a full energy spectrum, we discretise the energy
spectrum. Let $N_E$ be the number of sampling points and $E_j$ the
neutrino energies. The straightforward way to proceed is to define two
secondary grids for each value of the energy $E_j$.  Actually we do
not need so many secondary grids, we can form groups of energies for
which the relations given by Eqs.(\ref{taud}) and (\ref{taue})
hold. In our example for a neutrino energy spectrum with $1.5 \
\textrm{MeV} \le E \le 15 \ \textrm{MeV}$ a partition of the spectrum
can be, for instance, such as described in table~\ref{table:part},
which shows that only $6$ secondary grids are required.

\begin{table*}
  \caption{Spectrum partition for different radius.
    \label{table:part}
  }
  \begin{center}
    \begin{tabular}{cc}
      \hline \hline
      Energy ($\textrm{MeV}$) & Radius $R_{in}$ ($\textrm{km}$) \\
      $1,5$ &  $4.35$ \\
      $2.0$ &  $7.7$  \\
      $3 - 4$ & $11.0$ \\
      $5 - 7$ & $13.3$  \\
      $8 - 10$ & $13.8$  \\
      $11 - 15$ & $14.35$ \\
      \hline
\end{tabular}
\end{center}
\end{table*}

\begin{figure}
  \centering
  \includegraphics[width=0.5\textwidth]{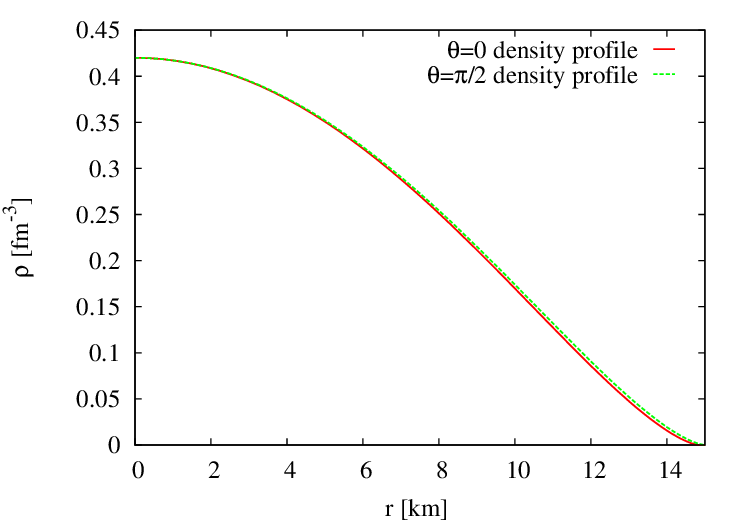}
\caption{Neutron star barionic density profile, equatorial and polar density.}
\label{fig:rho_r}
\end{figure}

Here we present the results: The matter density distribution at the
pole and at the equator of a rotating star with a rotation frequency
of 637 Hz is shown in Fig.~\ref{fig:rho_r}. In what follows, we shall
use this rotation frequency.

\begin{figure}
  \centering
  \includegraphics[width=0.5\textwidth]{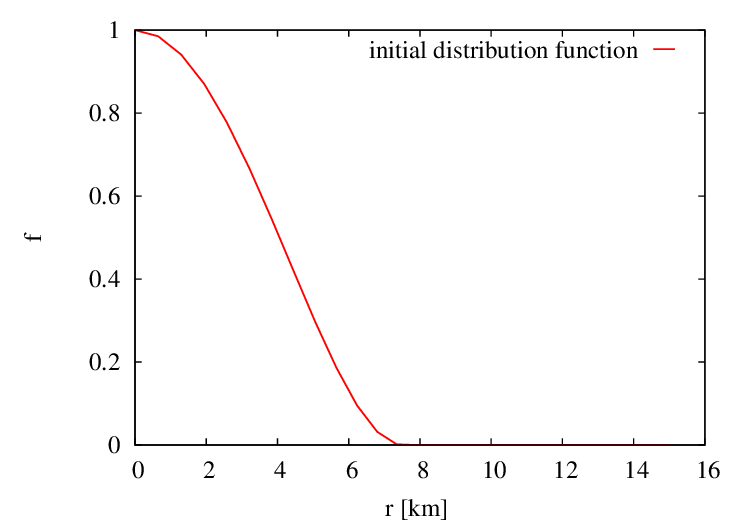}
  \caption{Initial conditions: Neutrino distribution function at time
    $t=0$ This initial distribution is the same for the three energies
    taken into account ($1.6 \ \textrm{MeV}$, $2 \ \textrm{MeV}$, $5 \
    \textrm{MeV}$)}
  \label{fig:f_r}
\end{figure}

\begin{figure}
  \centering
  \includegraphics[width=0.5\textwidth]{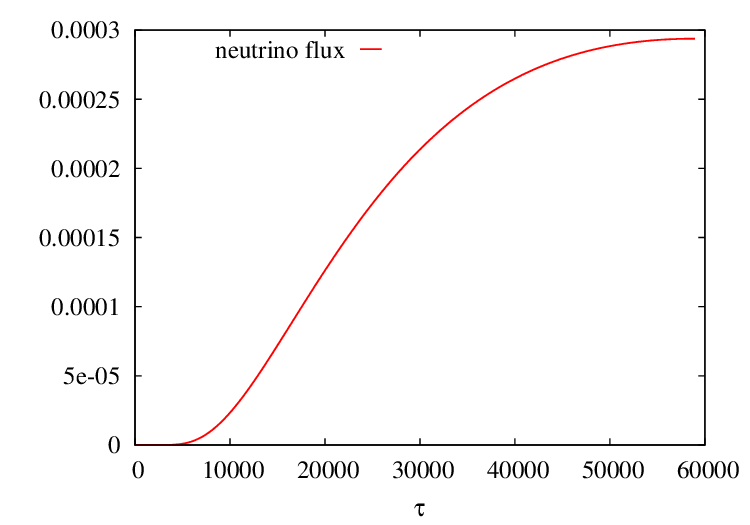}
  \caption{Neutrino flux at the surface of the star, with a neutrino
    energy of $2 \ \textrm{MeV}$. The $x$ axis
    represents the time in free fly unit $\tau = R/c$}
  \label{fig:fl_t}
\end{figure}

Figure~\ref{fig:f_r} shows the initial neutrino distribution function
at time $t=0$. This distribution depends only on $r$. The code runs
until the flux at the surface of the star reaches its maximum at the
time $t=T$ (See Fig.~\ref{fig:fl_t}).

\begin{figure}[h]
\begin{tabular}{ccc}
  \includegraphics[width=0.5\textwidth]{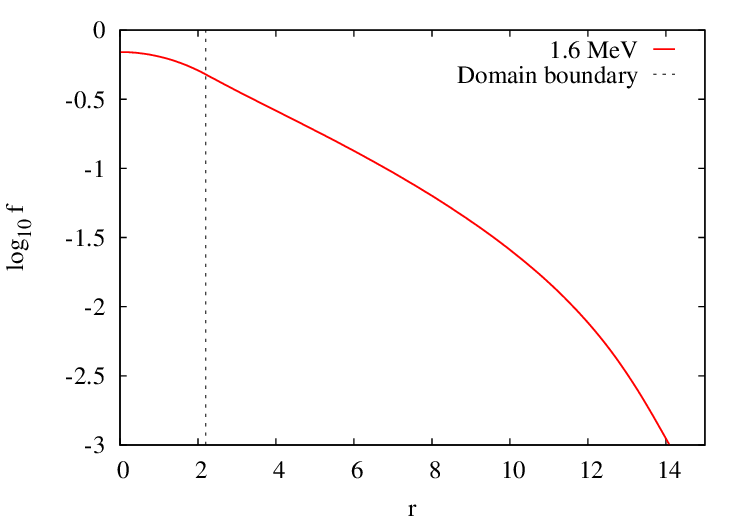}  \\

  \includegraphics[width=0.5\textwidth]{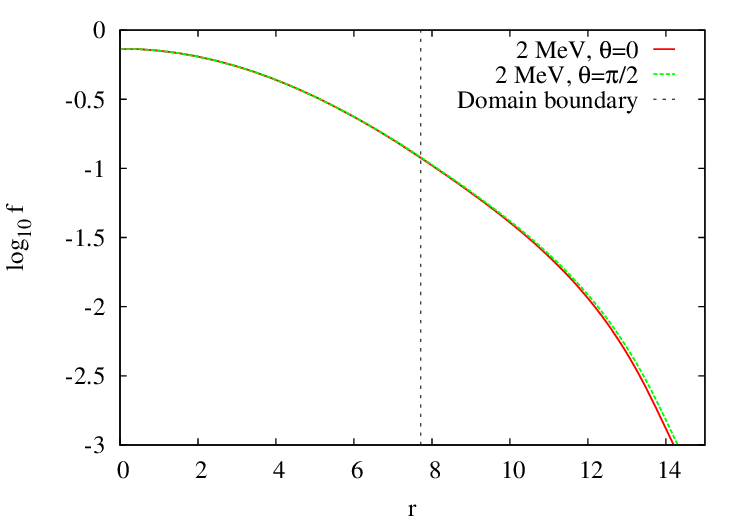} \\

  \includegraphics[width=0.5\textwidth]{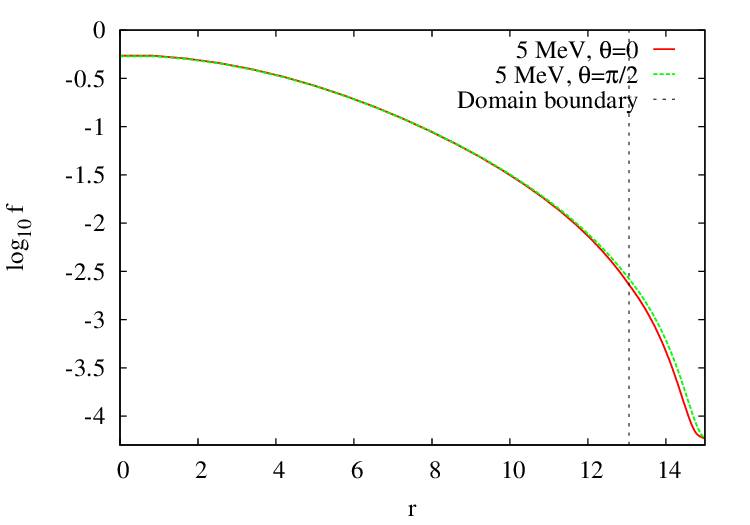}
\end{tabular}
\caption{$1.6 \ \textrm{MeV}$, $2 \ \textrm{MeV}$ and $5 \
  \textrm{MeV}$ neutrino distribution for $ \theta=0$ and $\theta=
  \pi/2$ at time $t=T$ (in the $1.6 \ \textrm{MeV}$, only one is
  represented because both $ \theta=0$ and $\theta= \pi/2$ are
  indistinguishable). The separation on the $x$ axis marks the
  boundary of the two grids (see text).}
\label{fig:logf_r_456}
\end{figure}

Figures~\ref{fig:logf_r_456} show the neutrino distribution function
for $\theta=0$ and $\theta=\pi/2$ averaged on $\Theta$ and $\Phi$ at
energies respectively of $1.6 \ \textrm{MeV}$, $2 \ \textrm{MeV}$ and
$ 5 \ \textrm{MeV}$. The star on the $r=R_{in}$ axis indicates the
separation of the two grids. The optical depth at $r=R_{in} $ was
chosen to be $\sigma n =5 \ \textrm{km}^{-1}$. Note the good matching
of the two solutions.

\begin{figure}
  \centering
  \includegraphics[width=0.5\textwidth]{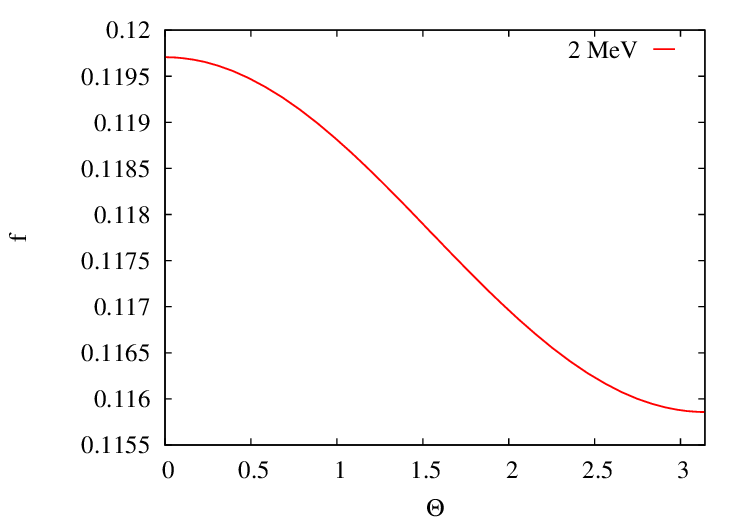}
  \caption{$2 \ \textrm{MeV} $ averaged neutrino distribution function as a
    function of $\Theta$ at $r=R_*$ at time $t=T$}
  \label{fig:f_T_7}
\end{figure}

Figure~\ref{fig:f_T_7} shows the neutrino distribution function
averaged on $\theta$ and $\Phi$ at the grids separation point in the
case of a neutrino energy $E = 2 \ \textrm{MeV}$.  Note that that the
function is very smooth across the axis $\Theta= \pi/2$

\begin{figure}
  \centering
  \includegraphics[width=0.5\textwidth]{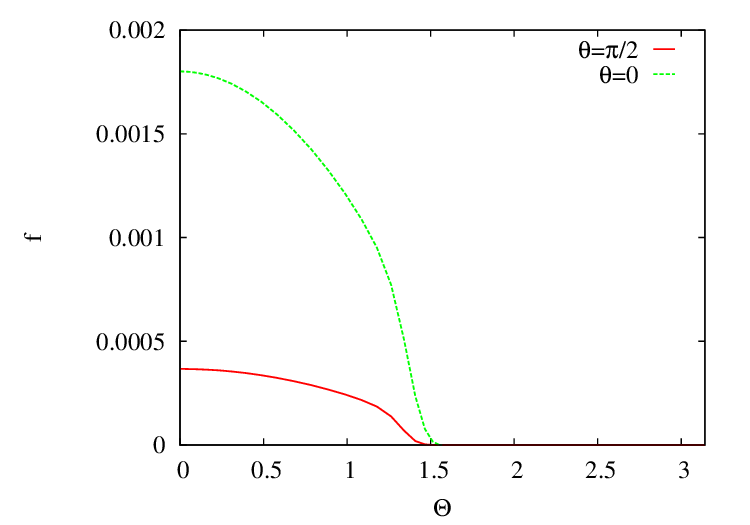}
  \caption{$2 \ \textrm{MeV} $ averaged neutrino distribution function as a
    function of $\Theta$ at the surface of the star}
  \label{fig:f_T_8}
\end{figure}

Figure~\ref{fig:f_T_8} shows shows the neutrino distribution function
$f(R,\theta,\Theta,\Phi,2 \ \textrm{MeV})$ averaged on $\theta$ and
$\Phi$ at the surface of the star. Note that the boundary conditions
(outgoing flux) is exactly fulfilled.

\begin{figure}
  \centering
  \includegraphics[width=0.5\textwidth]{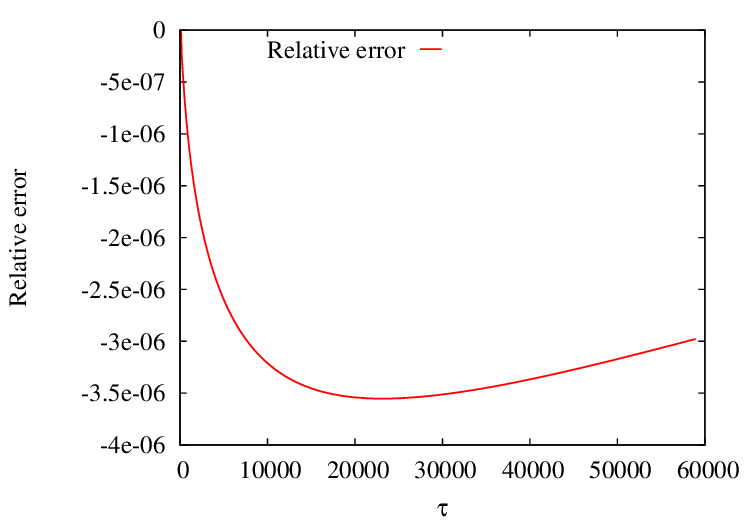}
  \caption{$2 \ \textrm{MeV}$ Relative error on conservation number of
    neutrino.}
  \label{fig:err_t}
\end{figure}

Fig.~\ref{fig:err_t} shows the neutrino conservation relative error as
a function of time (with a neutrino energy $E = 2 \ \textrm{MeV}$).

\vspace{1em}

The diffusion approximation holds only when the opacity $\sigma \, n
\; \to \, \infty $.  In order to estimate the relative error as
function of the thickness $\sigma \, n$, the error $\epsilon $ is
defined by

\be
\label{epsilon} \epsilon=
\frac{1}{P_0} \left( \sum_{l=2}^6 P_l^2\right)^{\frac{1}{2}}~,
\ee

where

\be
\label{P0l}
P_l=\int_0^{2\pi} d \Phi \int_0^{\pi} \, \sin
\Theta \, d \Theta \int_0^\pi \, P^0_l(\Theta)
F(R_{in},\theta,\Theta,\Phi,T) \,\sin \theta \, d \theta,
\ee 

where, again, $T$ is the time at the end of the run and
$P_l^0(\Theta)$ are the Legendre polynomials.

\begin{figure}
  \centering
  \includegraphics[width=0.5\textwidth]{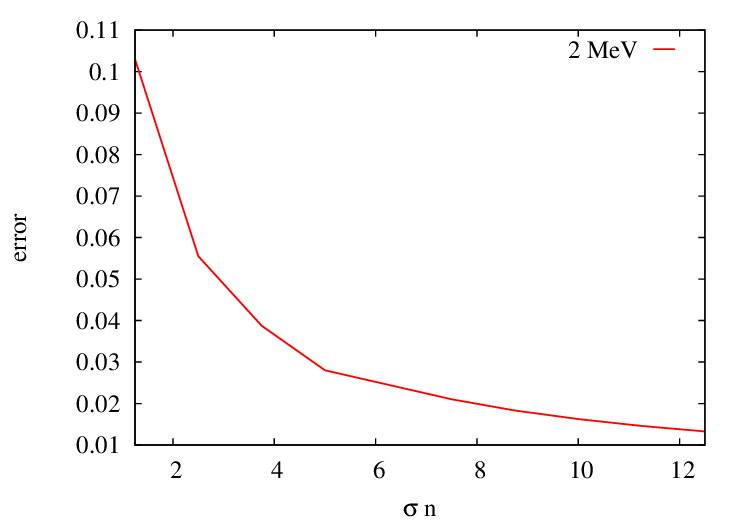}
  \caption{$2 \ \textrm{MeV}$ matching relative error. (See text)}
  \label{fig:err_thick}
\end{figure}

Fig.~\ref{fig:err_thick} shows the dependence of the error
$\epsilon$ on the optical thickness $\sigma \,n$. When moments
$P_{>2}=0$, the error vanishes to a good approximation.

Analogous errors are found for different energies.  It seems that an
optical thickness $n \sigma \sim 5 \ \textrm{km}^{-1}$ at the
grids separation is a good compromise.

\subsection{Convergence}

\begin{figure}
  \centering
  \includegraphics[width=0.5\textwidth]{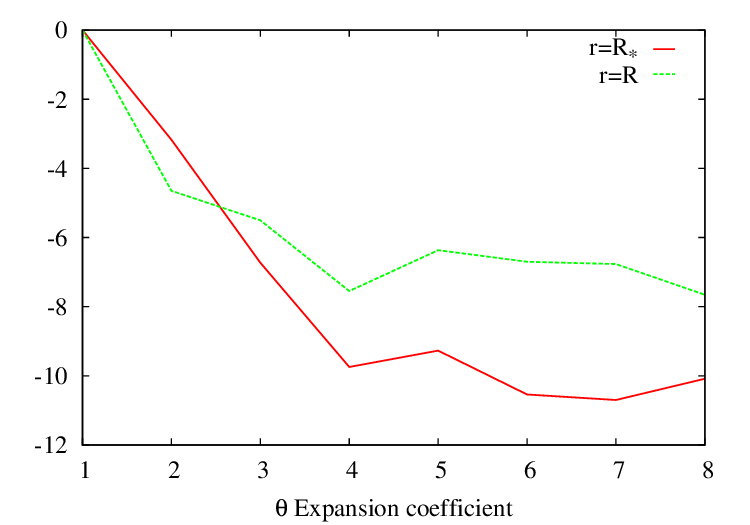}
  \caption{$2 \ \textrm{MeV}$ logarithm of averaged amplitude of
    $\theta$ coefficients expansion divided by the first coefficient,
    at $r=R_*$ (lower plot) and at
    $r=R$ (upper plot)}
  \label{fig:theta_coeffs}
\end{figure}

An efficient test to check the accuracy of the code consists in
studying the behavior of the amplitude of Chebyschev-Fourier
coefficients as a function of their order. Fig.~\ref{fig:theta_coeffs}
shows the behaviour of the Chebyschev normed coefficients of the
expansion in $\theta$ of the averaged solution.

\be
\label{Gtheta}
G_{\theta}(r,
\theta)=\int_0^{2\pi} d \Phi \int_0^{\pi} F(r,\theta,\Theta,\Phi,T) \,
\sin \, \Theta \, d \Theta.
\ee

For $r=R_{in}$ and $r=R$ we see that there exists a break of the slope
of the coefficients when the amplitude of the coefficients is $\sim
10^{-8} $. This behavior is due the fact that the matter density
derivative with respect to $\theta$ is discontinuous close to the surface
of the star (See Fig 1).

Note that only 8 coefficients are required to reach an accuracy of $
10^{-5}$. Here the number of coefficients is $17$, but the odd
coefficients vanish because of the equatorial symmetry of the problem.

\begin{figure}
  \centering
  \includegraphics[width=0.5\textwidth]{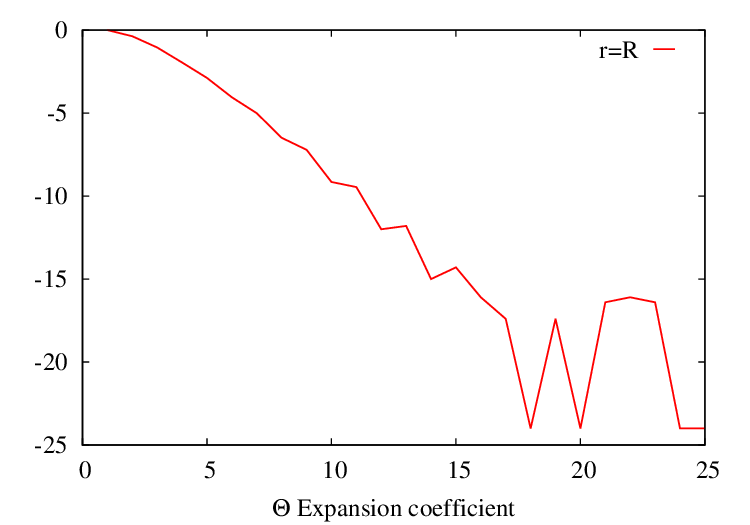}
  \caption{$2 \ \textrm{MeV}$ logarithm of averaged amplitude of
    $\Theta$ Tchebitchev expansion coefficients, divided by the first
    coefficient, at $r=R$}
  \label{fig:cap_Theta_coeffs}
\end{figure}

Fig.~\ref{fig:cap_Theta_coeffs} shows the Chebyschev normed
coefficients of the averaged functions $G_\Theta(r)$

\be
\label{GTheta}
G_{\Theta}(r,\Theta,T) =\int_0^{2 \pi} d \, \Phi \int_0^{\pi} 
F(r,\theta,\Theta, \Phi ,t)\sin \theta \, d \theta.
\ee

\begin{figure}
  \centering
  \includegraphics[width=0.5\textwidth]{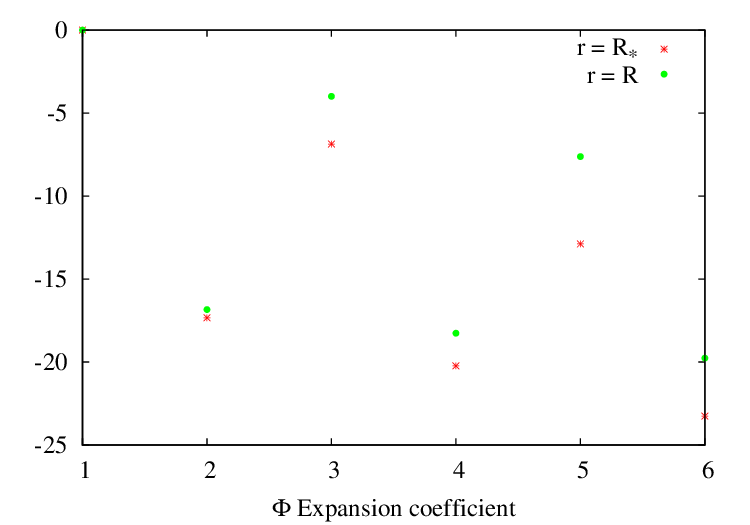}
  \caption{$2 \ \textrm{MeV}$ logarithm of averaged amplitude of
    $\Phi$ Fourier expansion coefficients divided by the first
    coefficient at $r=R_*$ (lower plot) and at $r=R$}
  \label{fig:Phi_coeffs}
\end{figure}

Analogously, Fig.~\ref{fig:Phi_coeffs} shows the behaviour of the $
\Phi$ coefficients.  Note that all the coefficients vanish, as
expected, exponentially when their number increase.

%

\subsection{Fast rotating star}

In the case of a fast rigidly rotating star, the centrifugal force
strongly deforms the surface of the star. The deformation generates a
derivative discontinuity of the matter density in the spherical grid,
where the steady state configuration of the star is computed.
Remember that spectral methods loose their efficiency when
discontinuities appear.  In order to overcome this difficulty we
propose the adoption the method used in computing
the steady state configuration of a  fast rotating star.

We make a coordinate transformation

\be
\label{rtransf} r^{'} =r + r^3
f(\theta)~,
\ee 

where $f(\theta) $ determines the surface of the star. The Liouville
operator $ {\cal L}_{sph} $ is slightly modified but there is no
change in the outlined procedure.

The problem is that the plasma is not at rest and violates the
validity of the hypothesis we have taken. To overcome this difficulty
we propose a reference frame transformation using a coordinate
transformation

\be
\label{phip}
\phi^{'}=\phi +\Omega t.
\ee

In this comoving frame, the matter is at the rest. We have to solve
the transfer equation modified by the metric terms generated by the
rotation (see e.g. \cite{debb_09} for a derivation of the equation
with the metric terms taken into account).

\vspace{1em}

We conclude this section by describing a strategy to compute the
cooling of a neutron star. The numerical problem lies in the presence
of two characteristic times, $\tau_1$ and $\tau_2$. $\tau_1$ is the
neutrino propagation time within the star (of the order of a few
milliseconds, see Fig 6). The second one $\tau_2$ is the
characteristic time given by the heat capacity of the star and the
energy flux. $\tau_2$ is of the order of years. In order to overcome
this difficulty, we proceed with a two times technique.
 
Consider a NS at time $t=0$ with a null neutrino distribution. Let
$T(r,0)$ be its temperature profile. By tacking the density and
temperature profile fixed, let the neutrino density $F(r,\theta,
\Phi,t))$ relax towards a steady state regime. (Of course the
neutrino emission and absorption coefficients are taken into
account). The first step can be time consuming, if the initial
neutrino distribution is far from the steady state one. Once a steady
state regime is reached, by using the neutrino flux we can compute the
new temperature distribution of the star, the flux being considered
frozen. With the new frozen temperature we re-compute the new neutrino
distribution. The number of time steps required to reach the new
steady sate neutrino distribution function is much shorter than the
previous one because we start from a distribution close to the relaxed
one.

\section{Conclusion}

The aim of this numerical work was to assess a ``proof of principle''
for the treatment of the full transport equation in 6D spherical
coordinates in a single core processor, in reasonable physical and
computational situations, and by means of the use of spectral methods
in phase space. We emphasize the fact that as far as we know, a
6-dimensional approach in spherical-like coordinates has never been
attempted before, and that consequently no comparison with existing
works can be made. A particular setting of the computational grid is
necessary for treating singular behaviour of some terms in the
Liouville operator.  Meaningful numerical results are obtained in a
very reasonable computational time, the most time consuming operation
being the computation of the Liouville operator. For problems where
Fokker-Planck-like approximations can not be used, it is possible that
the most consuming computation would be related to the collision term,
in which the thermal distribution of the plasma has to be taken in to
account.  (See Eq.(\ref{Liouvsph})). We believe that spectral methods
are suited to build an efficient algorithm for the treatment this
problem. We have also seen that the use of a fully spectral scheme in
treating the advection term can turn out to be useful if reduction of
the CPU time is a priority.  We believe that by using fairly
reasonable parallel computation on a small-scale cluster, one would be
able to perform multiple runs in physically relevant 6-dimensional
settings and in a really quick fashion. Although we are aware of the
fact that several ingredients are still to be added to the transport
description to use it in a physically relevant radiation hydrodynamics
code, our results support the fact that no fundamental technical
difficulty should arise in tackling those issues.

\begin{acknowledgements}
  We wish to thank Eric Gourgoulhon, J\'erome Novak and Micaela Oertel
  for the many fruitful discussions during the development of this
  method.  N.V. acknowledges the support of the Swiss National Science
  Foundation under the grant No PP002-106627/1, and of the French ANR
  Grant 06-2-134423 entitled "M\'ethodes math\'ematiques pour la
  relativit\'e g\'en\'erale". B.P. acknowledges the support of the
  SN2NS project ANR-10-BLAN-0503.
\end{acknowledgements}
\bibliographystyle{aa} 


\appendix
\section {Particle conservation in the transport equation} 
\label{app:partcons}

We will concentrate here on the pure coherent scattering case of
transport equation for 
photons, which writes:
\be\label{Cohersp}
\frac{1|}{c} \frac{\partial \, F}{\partial t} + {\cal L}_{sph} \,F
+\sigma_t F - \int_{4 \pi} \sigma_d(\vec{\omega} \cdot
\vec{\omega}^{'}) \sin \Theta^{'} d \Phi^{'}=0.
\ee
The above equation can be written, after multiplication by
the element volume of the space of phase $r^2 \sin \theta \sin \Theta $ as
\be\label{Cohercons}
\frac{1}{c} \frac{\partial}{\partial t} \left(F r^2 \sin \theta \sin
\Theta \right)
+\frac{\partial}{\partial r} \left(F r^2 \sin \theta \sin \Theta \cos
\Theta \right)
\ee
\be
+\frac{\partial}{\partial \theta} \left( F r \sin^2 \Theta \sin \theta
\cos \Phi \right)
-\frac{\partial}{\partial \Theta} \left( F r \sin^2 \Theta \sin \theta \right)
\ee
\be
+\frac{\partial}{\partial \phi} \left( F r \sin^2 \Theta \sin \Phi \right)
-\frac{\partial}{\partial \Phi} \left( F r \sin^2 \Theta \cos \theta
\sin \Phi \right)
\ee
\be
+r ^2 \sin \Theta \sin \theta \left( \sigma_t F - \int_{4 \pi}
 \sigma_d (\vec{\omega} \cdot \vec{\omega}^{'}) \,  F \sin \Theta^{'} d
  \Theta^{'} d \Phi^{`} \right) =0.
\ee  
After an integration on $r, \, \theta, \, \phi, \, \Theta, \, \Phi $,
and provided the detailed 
balance condition~\citep{Pomr1973} 
\be\label{DBC}
\sigma_d ( \, \gamma^{`} \to \gamma, \vec{\omega}^{'} \to \vec{\omega})=
\sigma_d (\, \gamma \to \gamma^{'}, \vec{\omega} \to \vec{\omega}^{'})   
\ee
holds, we obtain
\be\label{Cons}
\frac{\partial N}{\partial t}+ J_{R_{2}}-J_{R_{1}} =0,
\ee
where 
\be\label{Npho}
N= \int_{R_{1}}^{R_{2}}  r^2 \, dr \int_{4 \pi} \sin \theta \,d \theta 
\, d \phi \int_{4 \pi} \sin \Theta d \Phi F 
\ee
 is the number of particles
and
\begin{eqnarray}
\label{Flux}
J_{R_{2}} = R_{2}^2 \int_{4 \pi} \sin \theta \, d \, \theta \, d \phi \times \nonumber \\
 \int_{4 \pi} \sin \Theta \cos \Theta \, d \Theta \, d \Phi
F(t,R_{2},\theta,\phi,\Theta, \Phi)
\end{eqnarray}
is the flux of ingoing (outgoing) particles
into the surface $r=R_{2}$ of the spherical shell $ R_{1} \le r \le R_{2}$.
The same definition holds for $J_{R_{1}}$ with respect to the radius
$R_{1}$. 
This is a conservative form of the transport equation~(\ref{Cohersp}).

In the general case when energy dependence is taken in to
account, the photon number conservation is obtained after integration
on the energy $\gamma=h \nu/mc$. If induced scattering processes
  are also considered, 
The conservation equation Eq~(\ref{Cons}) is obtained
thanks to the detailed 
balance condition (\ref{DBC}).
\section{Enhanced spectral treatment in 2-D case: The conservative formulation}
\label{app:consform} 

We present a spherically symmetric version of an algorithm for
 a spectral treatment, amounting to the 2-D case
 $F(r,\Theta,t)$ 
for the distribution function, and restricted to coherent Thompson scattering interactions. This approach is useful to show how the
 a prospective full spectral treatment should be handled, and how its inherent
  difficulties can be overcome.
 As opposed to what we did previously, we shall now use the conservative form
 explicated in Eq.(\ref{Cohercons})) for the numerical representation of the distribution function:
We introduce the new function (see Appendix A)
\be\label{flux}
\hat{ F}(r,\Theta,t)=r^2 \sin \Theta F(r,\Theta,t)
\ee   
Using the previously defined value for the total Thompson cross section 
$ \sigma^{Tot}$ in photon scattering, the  2-D conservative form of the 
transfer equation reads, after integration on the $\Phi$ angle{\footnote{This formulation allows to us to check the
conservation law term by term: After an angular integration
on $\Theta $, the right-hand side and the second term on the left-hand side of
the Eq.(\ref{speccons}) vanish. The integrated first term on the left 
expresses then exactly the balance in radial flux.}: 
\begin{eqnarray}\label{speccons}
\frac{\partial \hat{F}}{\partial t}+ \cos \Theta \frac{\partial
  \hat{F}}{\partial r}
-\frac{1}{r} \left(\sin \Theta \frac{\partial \hat{F}}{\partial
  \Theta}+\cos \Theta \hat{F} \right)\nonumber \\
 =n(r) \sigma^{Tot} \nonumber  
\times \left[
-\hat{F}(r,\Theta,t)+\frac{3}{8}\sin \Theta \right. \nonumber \\ 
\left. \int_0^\pi \left( 1+(\cos \Theta
\cos \Theta{'} )^2 +\frac{1}{2} (\sin \Theta \sin \Theta{'} )^2 \right) 
\hat{F}(r,\Theta',t) d \Theta^{'} \right]
\end{eqnarray}
 We shall consider the same boundary problem that the one presented in the 5-D
hybrid case, namely $\hat{F}(r,\Theta, \, t=0)=0$ as initial
 value, 
$\hat{F}(R_1,\Theta,t)=\cos \Theta \sin \Theta$ for $ 0 \le \Theta
\le \pi/2 $ and $\hat{F}(R_2,\Theta,t) $ for $\pi/2 \le \Theta \le \pi $.
The time evolution will lead to a discontinuous solution in the radial
direction. Spectral methods are not suited to handle this
  kind of problem. In order to show that, consider the simple advection equation
 \be\label{advec1}
\frac{\partial \Phi}{\partial t}+C \frac{\partial \Phi}{\partial r}=0,
\ee
where $C > 0$ is a constant and $R_1 \le r \le R_2$, with the
initial data  $ \Phi(r,0)=0$ and the boundary condition
$\Phi(R_1,t)=1$. This problem is clearly analogous to ours, although
simpler; it is also well known that the solution is an Heaviside 
function $\Phi(r,t)=\Theta(r-Ct)$ propagating in our
setting from the inner radius $R_1 $ to the outer radius $R_2$ at
velocity $C$. We expect to obtain a solution to our problem with similar properties.
As we have already said, spectral methods are in general not well suited to treat discontinuous solutions,
except if some algorithm is used to smear out the solution
~\citep{Gott1977}}. In particular, 
it is possible to introduce viscosity in the spectral scheme, 
which can be partly treated in the coefficient
space~\citep{Bona1985}. 
However, such a scheme would have severe effects on conservation laws in our case. 
In what follows, we shall show an algorithm which attempts to overcome such difficulties. 
 
A classical first order implicit time discretisation to the simple
advection 
problem above (the so-called Euler method) leads to
\be\label{advec2}
\Phi^{j+1}(r)=\Phi^j(r) -C \Delta t \frac{\partial  \Phi^{j+1}}{\partial r},  
\ee
where $\Phi^j $ is the value of the solution at time $t=j
\Delta t$. and $\Delta t $ is elementary time interval.
We solve the above equation by making an expansion in Chebyshev
polynomials, imposing boundary values using a classical Tau approach~\citep{Gott1977} and with different values of the parameter $\Delta t$.\footnote{The matrix of
differential operator in the Chebychev basis can be reduced easily
thanks a linear combination of the lines, to a a tridiagonal matrix. 
This leads to a considerable speedup of the algorithm.}

Fig.~B.1. shows the numerical and analytical solutions obtained
with $N_r=65$ points in the propagation direction and $\Delta \,t=3(R_2-R_1)/(C
\, N_r^2)$ 
(this corresponds to 3 times the maximal value satisfying the stability Courant condition for an {\it
  explicit} numerical scheme).
As expected, we observe strong oscillations in the solution due to the
Gibbs phenomenon occurring at the solution discontinuity. Note
that the propagation
velocity for the solution seems however to be empirically correct.

 Numerical and analytical solutions using the much bigger value
 $\Delta t= 24(R_2-R_1)/(C N_r^2)$ 
are displayed in Fig.~B.1. While oscillations have disappeared, the
 numerical solution is spread out (the numerical propagation velocity being still correct).
\begin{figure*}
\begin{tabular}{cc}
\noindent \includegraphics[width=0.43\textwidth]{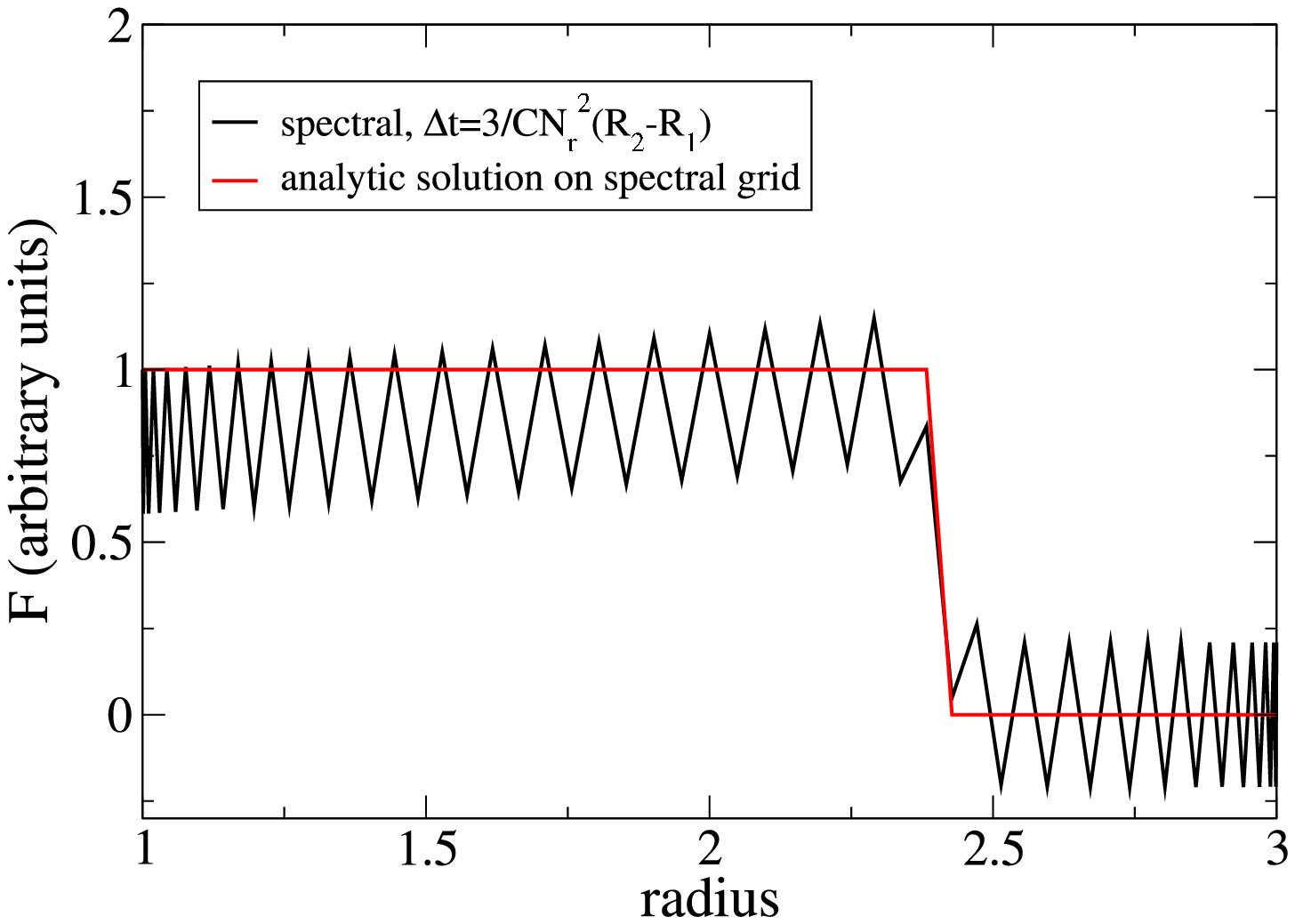}
&
\includegraphics[width=0.43\textwidth]{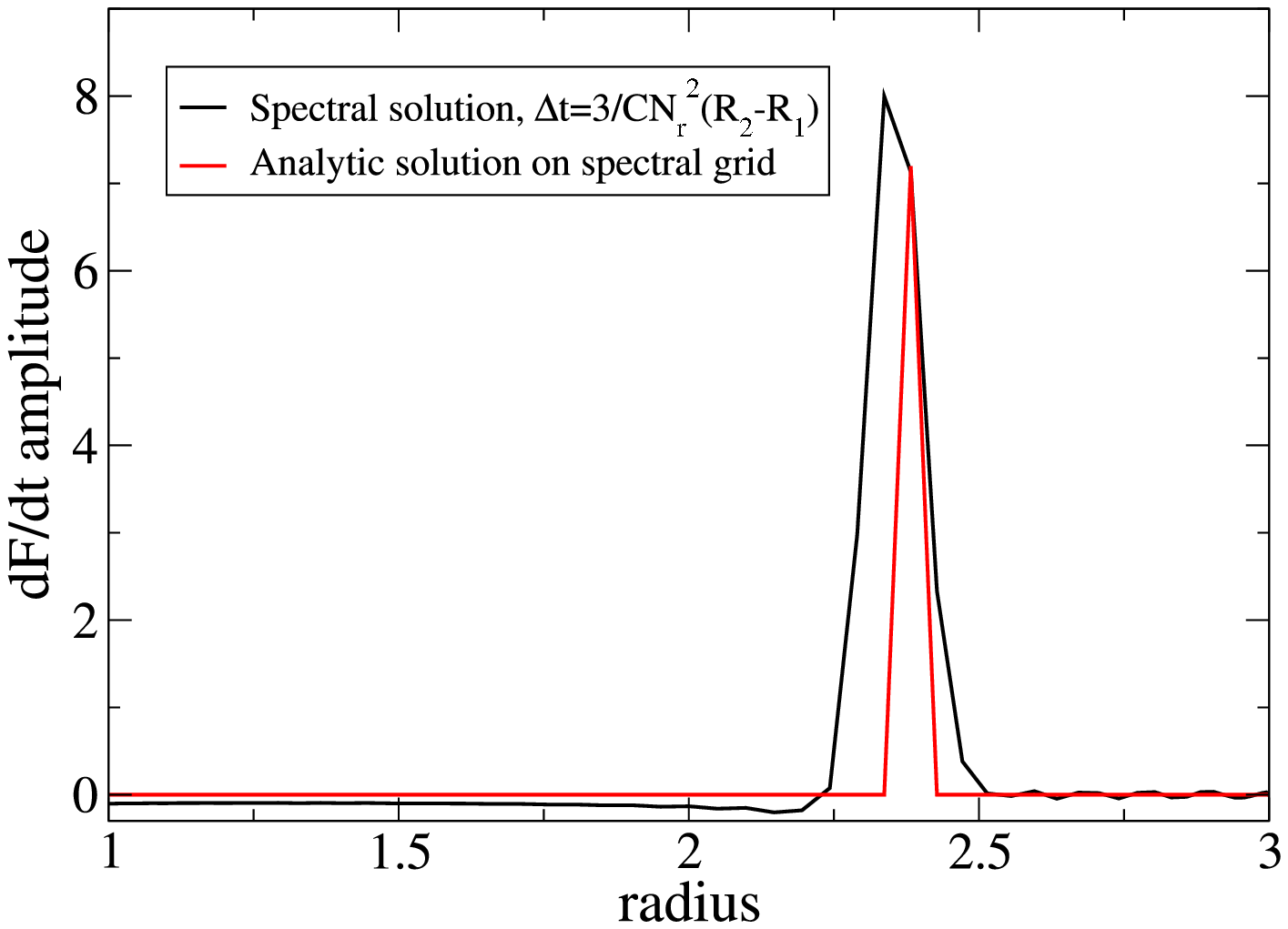} \\
\noindent \includegraphics[width=0.43\textwidth]{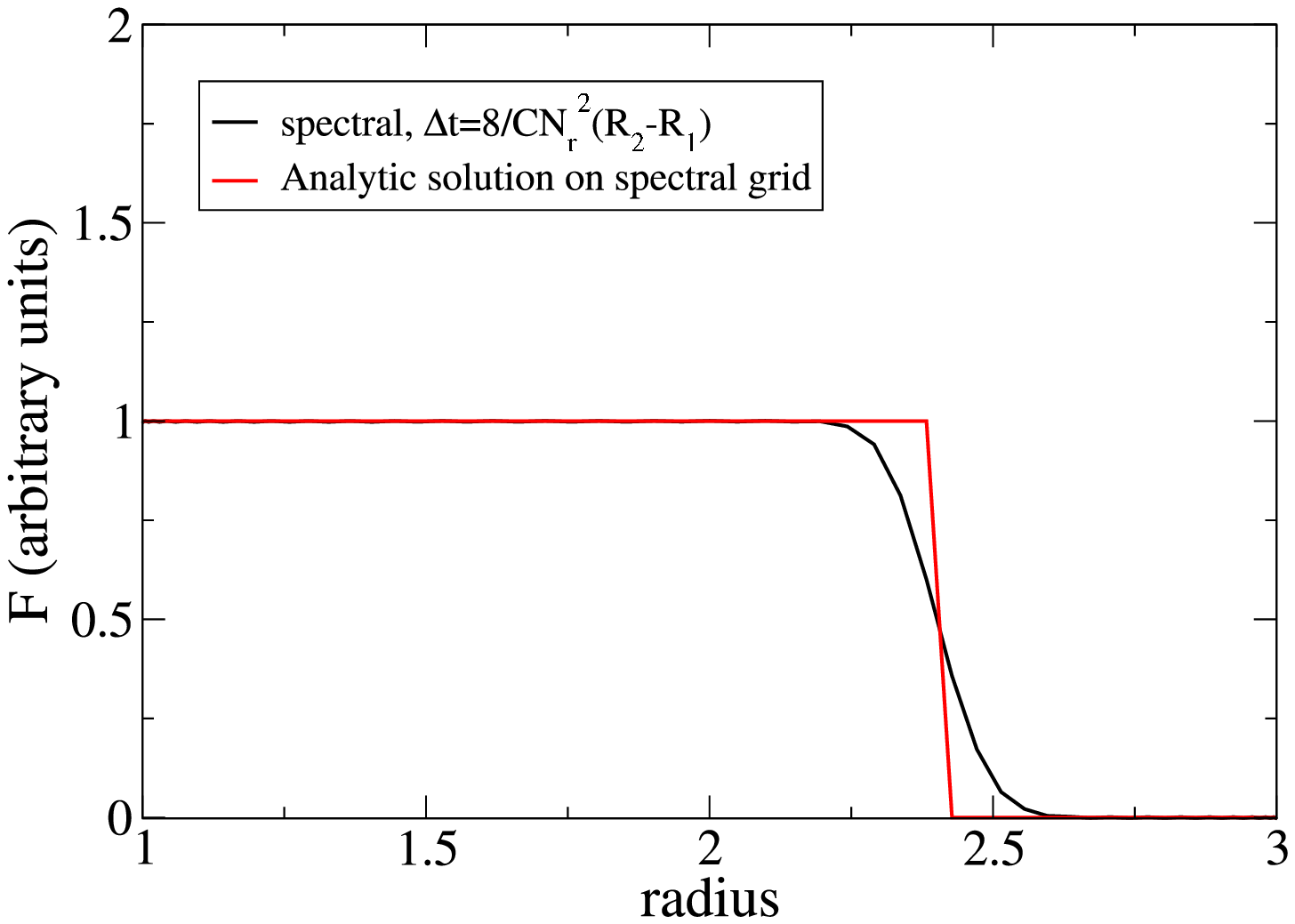}
&
\includegraphics[width=0.43\textwidth]{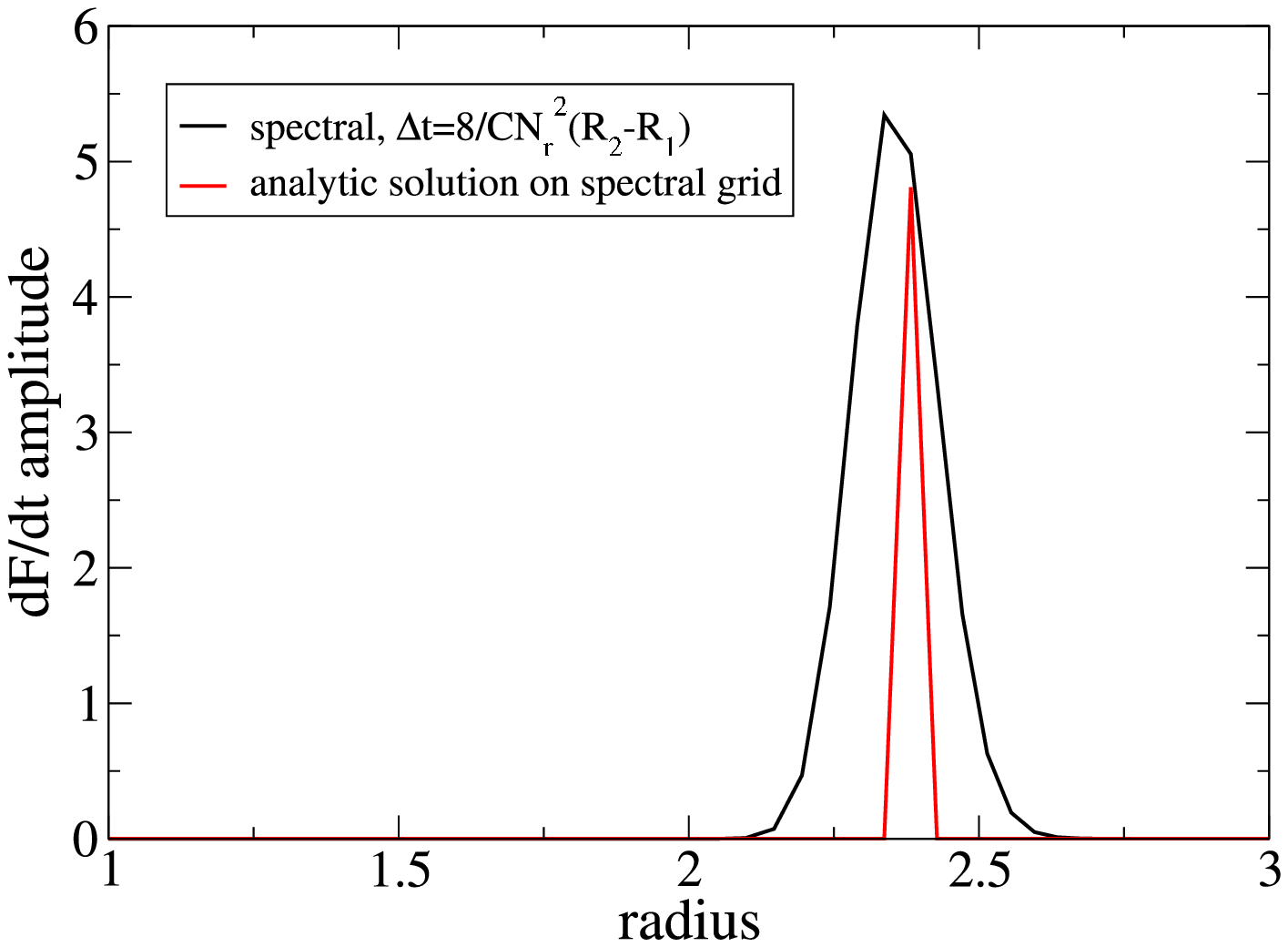}\\
\noindent \includegraphics[width=0.43\textwidth]{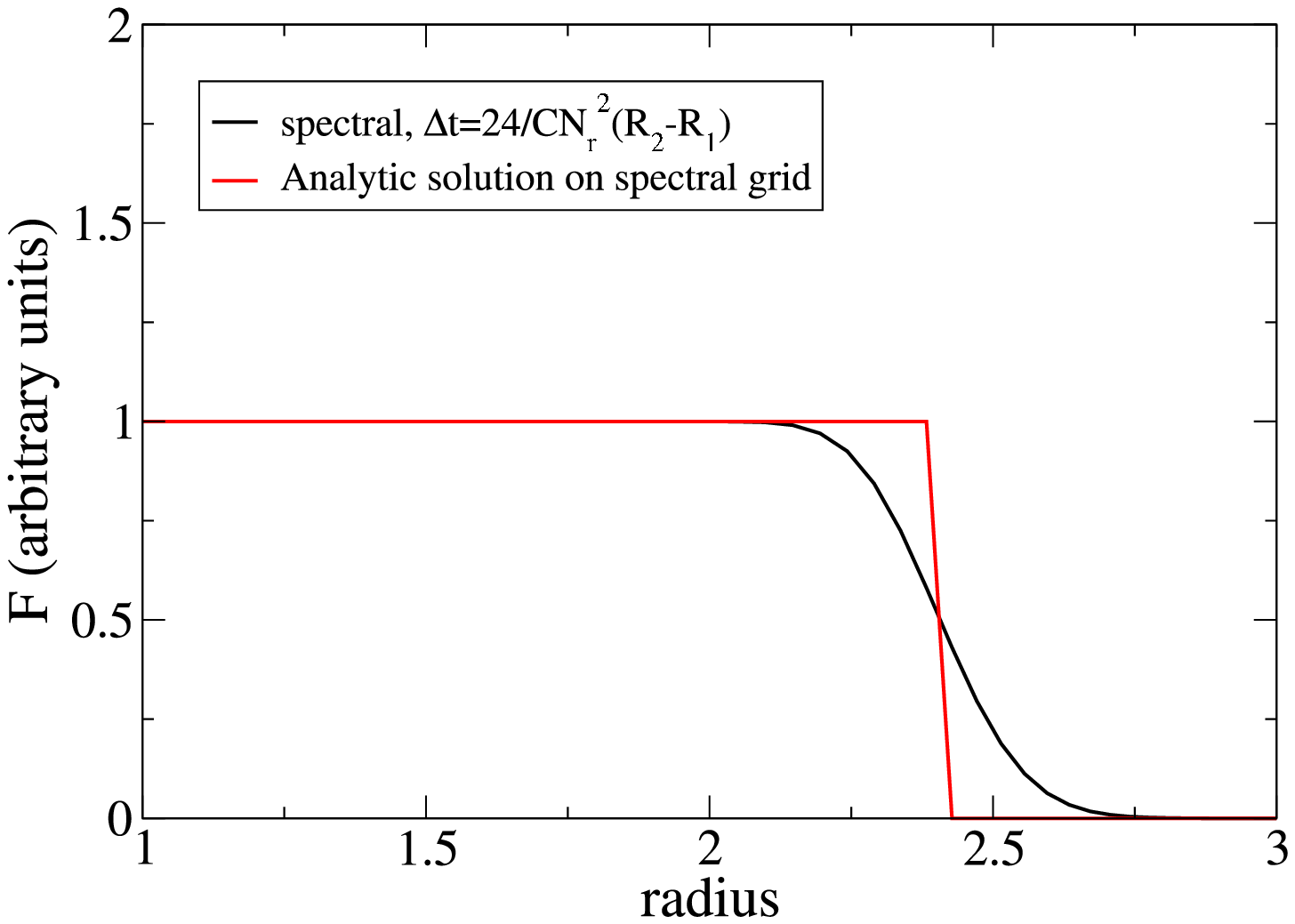}
&
\includegraphics[width=0.43\textwidth]{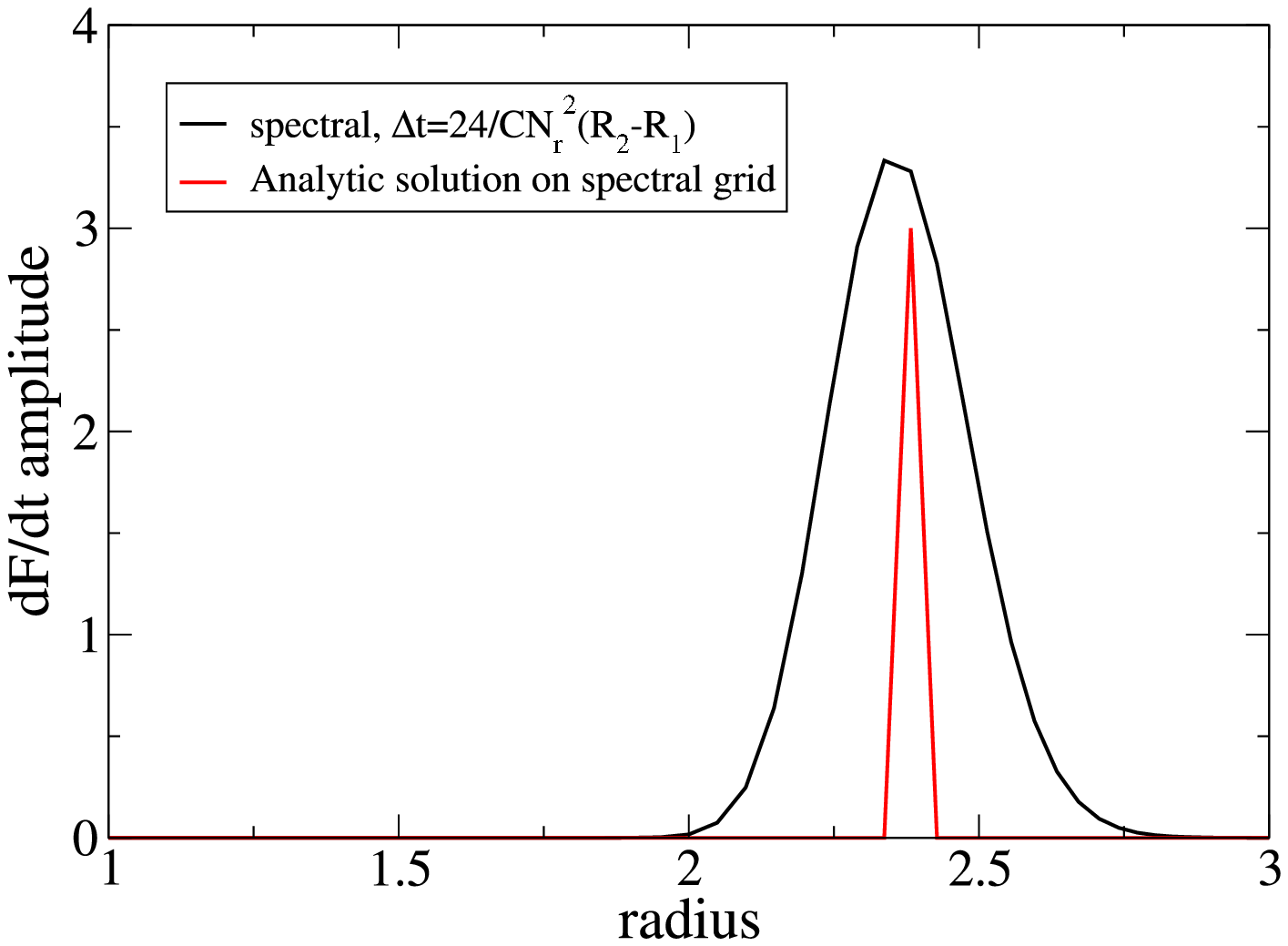}\\
\noindent \includegraphics[width=0.43\textwidth]{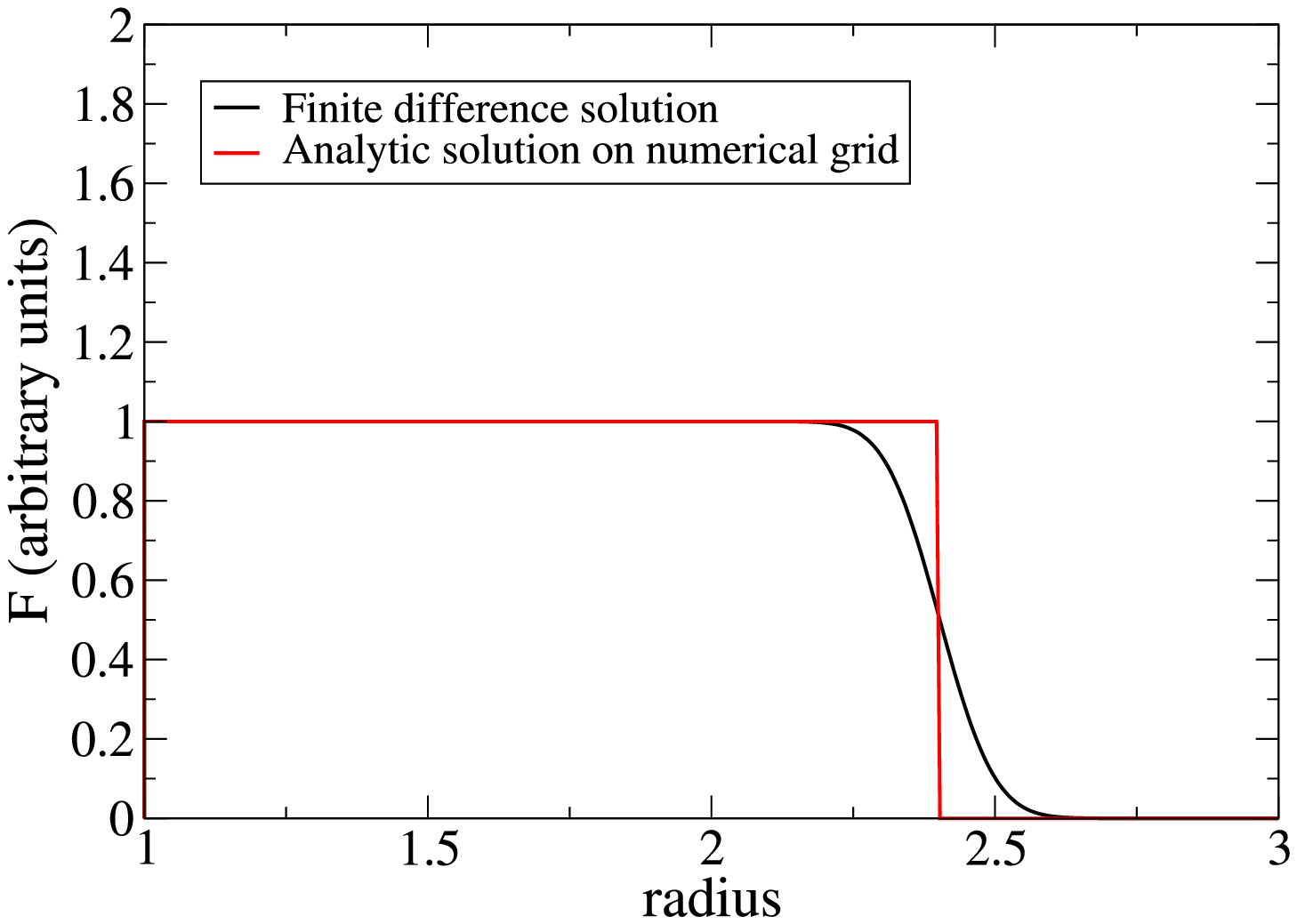}
&
\includegraphics[width=0.43\textwidth]{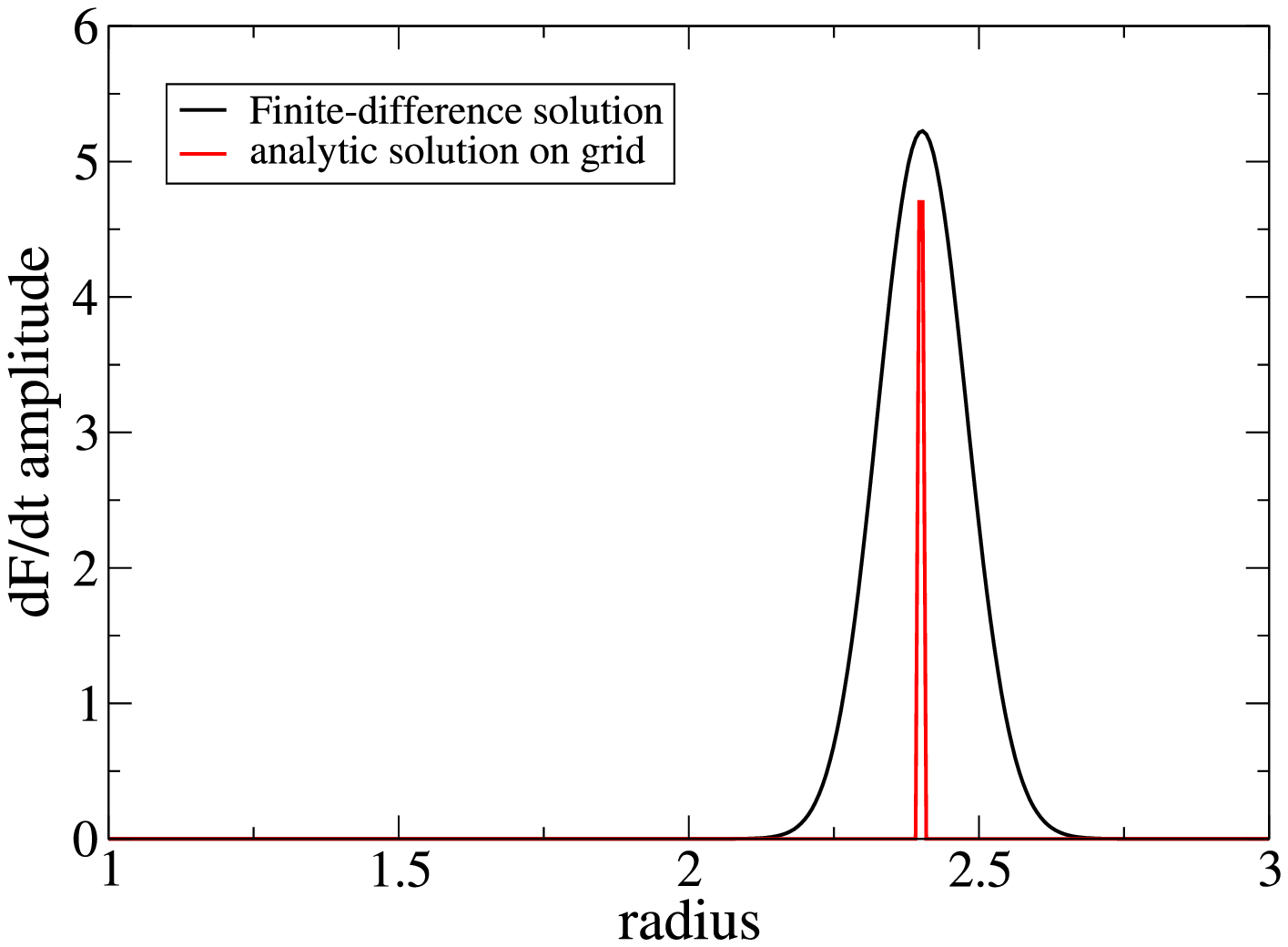}
\end{tabular}
\label{f:spvsFD}
\caption{Resolution of the advection problem for a shock profile
  using spectral methods. The left panels show the radial profile for  different 
choices of the time step, along with the analytical solution (interpolated 
to the spectral grid) and a finite-difference solution (lower left panel). In the  spectral solver, 
the number of grid points is $N_{r}=65$, whereas in the
  finite-difference solver, 
the choice $N_{r}=325$ is made. The right panel shows a comparison 
for the values of the function derivative in the spectral cases and the
  finite difference one. The amplitude of the derivative is a
  good indicator of the diffusivity of the numerical scheme. Note that
  the derivative of spectral solution is larger than the one of the
  finite difference scheme.
The value C=1 is here chosen.}
\end{figure*}
We have found experimentally that a value of about
\be\label{deltopt}
\Delta \, t_{opt}= \frac{8}{C N_r^2}(R_2-R_1) 
\ee
gives the best results, as shown in Fig.~B.1. The above
value seems quite 
independent on the number of spectral radial points: It holds for
$17\le N_r \le 257$ in this particular problem. On the contrary, we
have observed that a 
second order scheme of the type:
\be\label{advec3}
\Phi^{j+1}+\frac{C \Delta t}{2} \frac{\partial  \Phi^{j+1}}{\partial r} 
=\Phi^j -\frac{C \Delta t}{2} \frac{\partial \Phi^j}{\partial r}
\ee
 leads to an incorrect propagation velocity. \\
In order to compare the results obtained with the first order
finite difference scheme and the spectral one, we plot the values for 
the derivative $\partial_r \Phi$ (see Fig.~B.1. We recover similar results for the amplitude
 of derivatives when the grid point number ratio between the finite
differences scheme and the spectral one is about five (330
points versus 65). 
The size ratio mentioned in the introduction seems then to be verified.
\begin{figure}[h!]
\begin{center}
\includegraphics[width=0.5\textwidth]{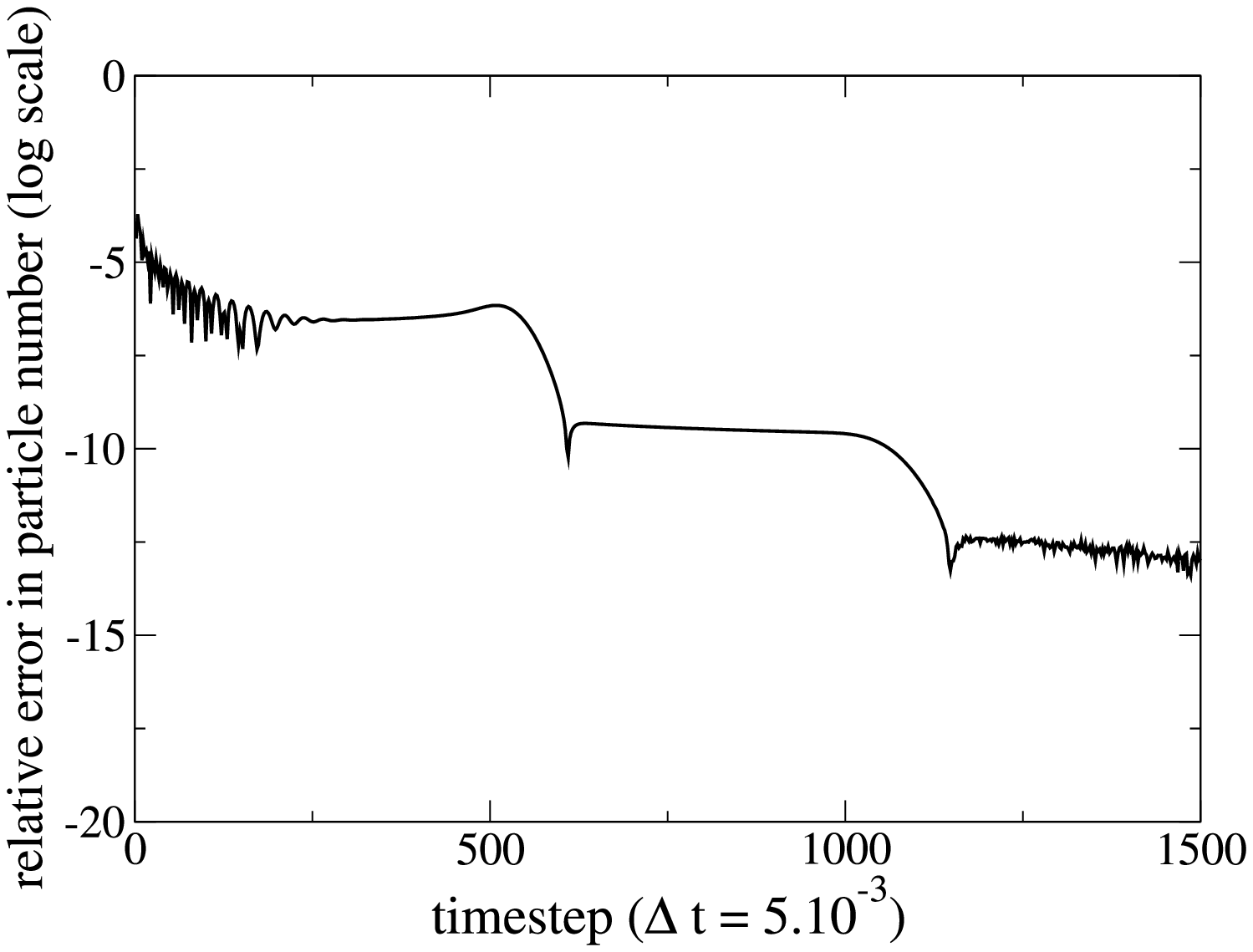}
\end{center}
\caption{Relative variation of particle number at a given time step,} 
\label{f:conssimple}
\end{figure}
Finally, Fig.~\ref{f:conssimple} shows the error on the 
conservation of particles for this scheme, namely the verification of the identity:
\be\label{photcons}
\frac{\partial}{\partial t} \int_{R_1}^{R_2} \Phi(r,t) dr +C \left(
\Phi(R_1,t)-\Phi(R_2,t) \right)=0. 
\ee

The above simple and fast algorithm seems indeed to be able to
handle correctly the discontinuities in the solution, keeping conservative features at the same time. It is tempting to apply it in solving the Eq.(\ref{speccons}).
Let us write this equation in the following effective way:
\be\label{specon2}
\hat{F}^{j+1}_{k}-\cos \Theta_k \frac{\partial
    \hat{F}_{k}^{j+1}}{\partial r} \Delta \,t=
\hat{F}^j_{k} +\Delta \,t S_{k}^{j+1/2}
\ee
 where $\hat{F}^{j}_{k}= \hat{F}(r,\Theta_{k},t_{j})$, $\Theta_k$ 
is the value of the discretised variable $\Theta$, and $S_{k}$ contains 
all differential terms on $\hat{F}$, taken on the same values, appearing in Eq.~(\ref{speccons}).
The coefficient in front of $\partial_r \hat{F}$ depends on the variable 
$\Theta$. As a consequence, we cannot {\it a priori} define a consistent optimal value 
$\Delta t_{opt}$ for every value of $\Theta$ as in Eq.(\ref{deltopt}).  
Moreover, we want to be free in choosing
the value of $\Delta t$, which will in general be constrained
by a Courant stability condition, either related to the transport equation itself or an adjacent hydrodynamic scheme.
We proceed then in the following way: consider first an time explicit version of Eq.~(\ref{specon2}):
\be\label{specon3}
\hat{F}^{j+1} =
\hat{F}^j_{k} +  \cos \Theta_k \frac{\partial
    \hat{F}^{j}_{k}}{\partial r} \Delta \,t + \Delta \,t S^{j}_{k}.
\ee

This can be viewed as a set of $N_{\Theta}$ equations, where terms are
evaluated for 
each value of the discretized angle $\Theta_{k}$. We define for each of those angles an optimal time step:
\be \label{e:dtopt}
\Delta t_{opt}^{k}= \frac{8}{cos(\Theta_{k})N_{r}^{2}} (R_2-R_1).
\ee

 If $\Delta t_{opt}^{k} \geq \Delta t$, we compute the variation
\be \label{e:fdtopt}
G_{k}^{j}(t)=\hat{F}_{k}(t_{j}+\Delta t_{opt}^{k})- \hat{F}_{k}(t_{j}) = \Delta t_{opt}^{k}\cos \Theta_k \frac{\partial
    \hat{F}_{k}(t_{j})}{\partial r}.
\ee

 A simple linear interpolation is then performed to obtain the updated value for $\hat{F}^{j+1}_{k}$: 
 \be\label{interp2}
\hat{F}(r, \Theta_{k}, t_{j}+\Delta t) = \hat{F}^{j+1}_{k}= \hat{F}^j_{k}+ G_{k}^{j} \, \cos \Theta_k
  \frac{\Delta \,t}{\Delta \, t_{opt}^{k}} + \Delta_{t}S^{j}_{k}.
\ee 
 
 The case $\Delta \,t > \Delta \, t_{opt}$ is treated by introducing
an intermediate time interval 
\be \label{geDelta}
 t_{int}^{k} = \Delta \,t /K \le
\Delta \, t_{opt}^{k},
\ee
 where $K$ is the smallest integer satisfying the above relation, and performing the numerical integration $K$ times per global time step.
  
 In the implicit setting of Eq.~(\ref{specon2}) and with  $\Delta t_{opt}^{k} \geq \Delta t$, we slightly correct the previous scheme by defining: 

\be
  G_{k,\epsilon}^{j}(t_{j})= \Delta t_{opt}^{k}\cos \Theta_k \frac{\partial
    \hat{F}_{k}(t_{j})}{\partial r} + \epsilon(\Delta t_{opt}^{k})^{2},
\ee
 where the real parameter $\epsilon$ is tuned in the algorithm so that the partial update

\be
{\hat{F}^{j+1*}_{k}}= \hat{F}^j_{k}+ G_{k, \epsilon}^{j} \, \cos \Theta_k
  \frac{\Delta \,t}{\Delta \, t_{opt}^{k}}
\ee
 satisfies exactly a radial flux balance for particle number. The update is completed by the implicit first-order step
\be
\hat{F}^{j+1}_{k} = \hat{F}^{j+1*}_{k} + \Delta t S^{j+1}_{k}.
\ee

The case $\Delta \,t > \Delta \, t_{opt}$ is also performed by splitting time updates as in Eq.~(\ref{geDelta}).
\begin{figure}[h]
\begin{center}
\includegraphics[width=0.5\textwidth]{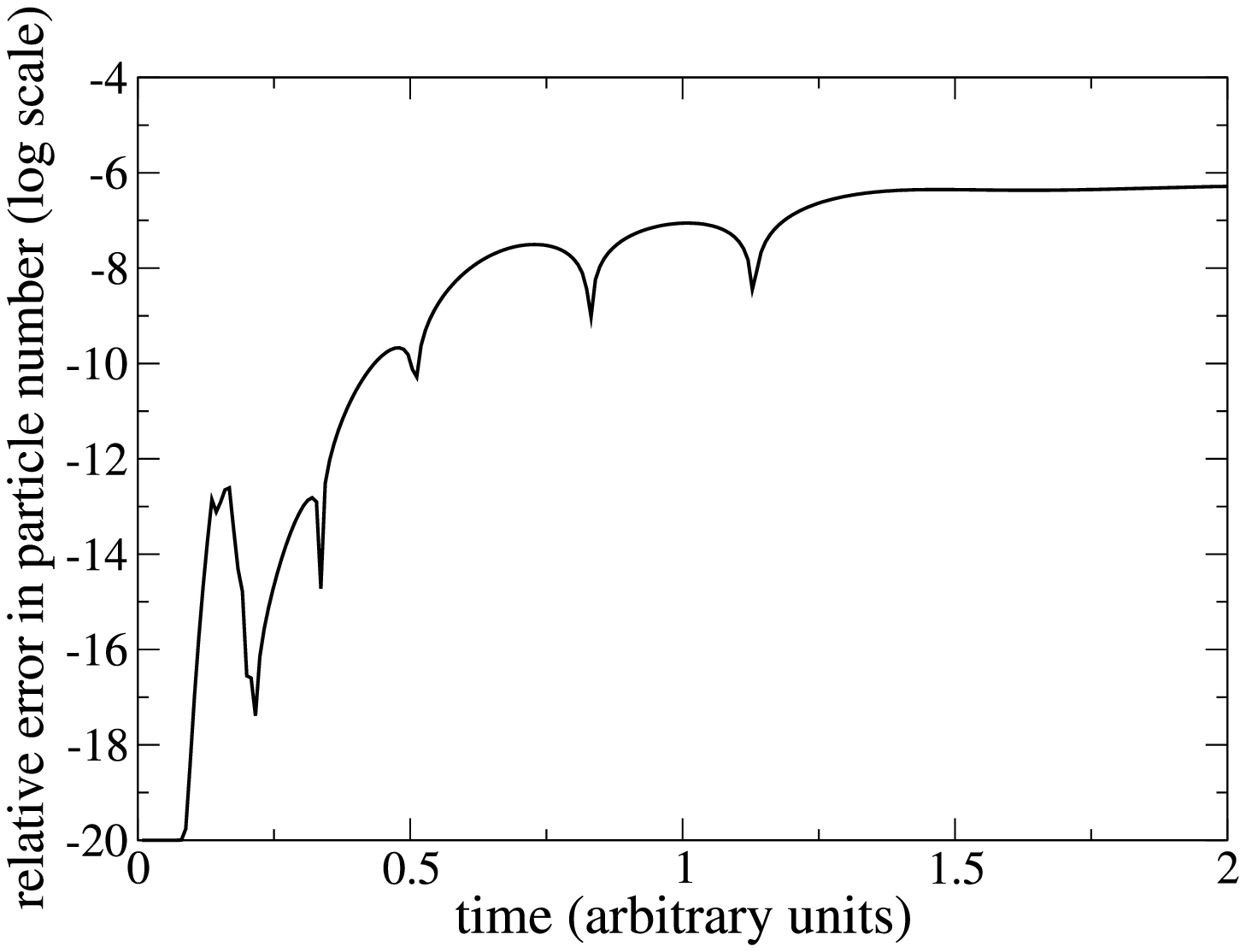}
\end{center}
\caption{Relative variation of particle number at a given time step, 
for the hybrid resolution of the 2D transport
equation~(\ref{speccons}).
The number of spectral points is $N_{r}=65$., with two finite
difference zones of 30 radial grid points at the edges.
 The error is mainly due to the differential term on $\Theta$. It
  decreases exponentially when the number of grid points in $\Theta$ increases.}
\label{f:constrans}
\end{figure}

In solving the  Eq.(\ref{speccons}) in the domain represented in
Fig.~\ref{f:Dom2d}, 
strong oscillations due to discontinuities may appear near the edges
of the interval, 
if the plasma density there does not vanish (see Sect. 5.2 and
Fig.~\ref{f:ADV}). 
A spectral resolution is in principle not able to handle these oscillations.
To overcome this problem, We have split the interval in 3
sub-intervals, two of them being close to the radial edges of the
main interval, and each outer sub-interval having a width of $0.1$ of
the main interval. The solution in the
sub intervals is computed with a finite difference scheme, using
$30$ grid points. The solution in the largest central interval
is computed with the spectral scheme presented just above, using $65$ spectral points and again
performing an expansion on a Chebychev polynomial basis. A Chebychev
polynomial expansion is also, as before, 
used to treat the $\Theta$ dependence. 

Results on Fig.~B.3 show the conservation of the number of
particles using this approach, with the settings of Sect. 5.2.  This
shows the validity of our scheme in the bulk and at domain boundaries,
and the accuracy of the conservative formulation in this 2D example.

Once known $\hat{F}=r^2 \sin \Theta F$, the distribution function $F$
is recovered easily by manipulating the coefficients in the spectral
decomposition; the division by  $\sin \Theta$ is nicely handled in the
  coefficient space, whereas the division by $r$ is performed in the
  configuration space. 

In this test, $25$ points are used in the $\Theta$ direction, and $(30+25+30)$ points in the radial direction for finite difference and spectral zones. The only Courant constraint for the timestep in the radial direction comes from the finite-difference zones, as an implicit spectral resolution in the radial direction is performed in the central domain. The timestep used is then $\Delta_t=5 \times 10^{-3} $.

In conclusion, treating the transport equation with a fully spectral
code is not straightforward.  In the above (hybrid) example in the
radial direction, the simulation requires a total of $125$ grid points
in $r$. As shown previously, in order to obtain the same accuracy with
a first order finite scheme, a rough number of $330$ grid points would
be necessary. This leads to a size ratio of $2.64$, two times less
than previously expected.

It is possible that the advantage of a spectral scheme reduces with
more sophisticated higher order finite differences schemes. However,
to match the performances of the presented approach, such a scheme
should be of order 2 or more, exhibit weak diffusivity and show no
oscillations due to discontinuities.

\section{The two moments approximation: Diffusion and  telegraph
  equation}
\label{app:diffeq}

The two moments approximation consists of making a spherical harmonics
expansion in $\Theta$ and $\Phi$ and neglecting all the moments higher
than one.  We obtain the well known diffusion equation and the
telegraph equation that can turn out to be interesting in the case of
fast time variability of the matter density.

Let us write the transfer equation for neutrinos in the general form

\begin{eqnarray}
\label{Transfdi}
\lefteqn{\frac{1}{c} \frac{\partial F}{\partial t} + \vec{\nabla} F \cdot
\vec{\omega} + n(r,\theta,\phi) +}\nonumber\\ 
&&\left[  4 \, \pi F(r,\theta,\phi,\Theta,\Phi,t) \right. \nonumber\\
&&\left. -\int_0^{2 \pi} d \, \Phi^{'}  \,
\int_0^{\pi} (a + b \,
\vec{\omega} \cdot \vec{\omega}^{'}) F(r,\theta,\phi,\Theta^{'}, \Phi^{'},t)
  \sin \Theta^{'} d \, \Theta^{'}\right]\nonumber\\
&&= S(r, \theta, \phi, t),
\end{eqnarray}

where $a(E) + b(E) \, \vec{\omega} \cdot \vec{\omega}^{'} $ is the
differential neutrino nucleon cross section $\sigma_D$,
$n(r,\theta,\phi)$ is the baryonic density, and $S(r,\theta,\phi,t)$
contains the source terms.  We consider only the two first moments:

\be
\label{Fmom0}
F(r,\theta,\phi,\Theta,\Phi,t)=
F^0(r,\theta,\phi,t)+3 \,\vec{F}^1(r,\theta,\phi,t) \cdot \vec{\omega}~.
\ee

By averaging over $\Theta$ and $\Phi$: $1/4 \pi \int_{\Omega} d \Omega
$ we obtain 

\be
\label{0mom} \frac{1}{c} \frac{\partial F^0}{\partial
  \, t}+\vec{\nabla} \cdot \vec{F}^{1} =S.
\ee

After multiplying $ F(r,\theta,\Theta,\Phi,t) $ by $\,
\vec{\omega}$, averaging over the solid angle gives

\be
\label{mom1}
\frac{1}{c}\frac{\partial \vec{F}^1}{\partial t}+ \frac{1}{3}
\vec{\nabla} F^0 + \tilde{\tau} \vec{F}^1=0,
\ee

where $\tilde{\tau}$ is the optical dept

\be
\label{optdep}
\tilde{\tau}= 4 \, \pi \left(a(E) - \frac{1}{3} b(E) \right)\,
n(r,\theta,\phi,t).
\ee

The {\it Eddington} (or diffusion) approximation consists in
neglecting the time derivative in Eq.(\ref{mom1}). In this case we
have the Fick law

\be
\vec{F}^1=-\frac{1}{3 n \, \tilde{\tau}}
\vec{\nabla} F^0~,
\ee
      
and the diffusion equation reads

\be
\label{dif}
\frac{1}{c} \frac{\partial F^0}{\partial t}
-\nabla_j \left( \frac{1}{\tilde{3 \, \tau}} \nabla^j F^0 \right)=S.
\ee

We propose to go further with the approximation. To do this, take the
time derivative of Eq.(\ref{0mom})

\be
\label{dt0mom}
\frac{1}{c^2}
\frac{\partial^2 F^0}{\partial t^2}+\frac{1}{c} \frac{\partial \,
  \nabla_j F^{1 \,j}}{\partial t} =\frac{1}{c}\frac{\partial \,
  S}{\partial t}~,
\ee

and take the divergence of Eq.(\ref{mom1})

\be
\label{divmom1}
\frac{1}{c} \frac{\partial \, \nabla_j F^{1, j}}{\partial t}
+\frac{1}{3} \left( \Delta F^0 + \left( \nabla_j \tilde{\tau} \right)
  F^{1 \, j} + \tilde{\tau} \nabla_j F^{1 \,j} \right) =0.
\ee

By replacing $\partial \, \nabla_j F^{1 \,j} / \partial t$ in
Eq.(\ref{dt0mom}) we obtain

\be
\label{dt2F0}
\frac{1}{c^2}
\frac{\partial^2 \, F^0}{\partial t^2}- \frac{1}{3} \left( \Delta F^0
  + \left( \nabla_j \, \tilde{\tau} \right) F^{1 \, j} +\tilde{\tau} \nabla_j F^{1
    \,j} \right) = \frac{1}{c} \frac{\partial S} {\partial t}~.
\ee

In the above equation the term $\nabla_j F^{1 \, j}$ comes from the
one obtained in Eq.(\ref{0mom}) and $(\nabla_j \tilde{\tau}) F^{1
  \,j}$ comes from Eq.(\ref{mom1}) in which the time derivative is
neglected.  In the end, we obtain the following telegraph equation

\be
\label{Teleraph}
\frac{1}{\tilde{\tau}} \left( \frac{3}{c^2}
    \frac{\partial^2 F^0}{\partial t^2}-\Delta F^0 \right)+\frac{1}{c}
  \frac{\partial F^0} {\partial t} - \nabla_j \,
  \left(\frac{1}{\tilde{\tau}} \right) \nabla^j F^0=S+ \frac{1}{c
    \tilde{\tau}} \frac{ \partial S}{\partial t}~.
\ee

When the time variations are weak, we recover the diffusion equation,
Eq.(\ref{dif}). The propagation velocity of the signal is $ \le c/
\sqrt{3} $ which is more satisfactory, in a relativistic context,
especially when the matter motion is close to the velocity of the
light, than the diffusion equation which gives an infinite propagation
velocity.  Note that if $\tilde{\tau}$ is constant the result is
exact.

\end{document}